\documentclass[12pt]{article}
\pdfoutput=1
\usepackage{amssymb, amsmath,amsfonts}
\usepackage{multirow}
\usepackage{mathrsfs}
\usepackage{array}
\usepackage{cite}
\usepackage{booktabs}
\usepackage{tikz}
\usepackage{pgfplots}
\usepackage{float}
\usepackage{subfigure, graphicx}
\usepackage[pdftex, bookmarks=true,colorlinks,linkcolor=red,urlcolor=blue,citecolor=blue]{hyperref}
\usepackage{cite}
\usepackage{slashed}
\usepackage{tabularx}

\usepackage{tikz}

\makeatletter\@addtoreset{equation}{section}\makeatother

\def\be{\begin{equation}}
\def\ee{\end{equation}}
\def\bea{\begin{eqnarray}}
\def\eea{\end{eqnarray}}
\newcommand{\comment}[1]{{\bf {\textcolor{blue}{ [#1]}}}}

\def\Dslash{\,\,{\raise.15ex\hbox{/}\mkern-12mu D}}
\def\Dbarslash{\,\,{\raise.15ex\hbox{/}\mkern-12mu {\bar D}}}
\def\delslash{\,\,{\raise.15ex\hbox{/}\mkern-9mu \partial}}
\def\delbarslash{\,\,{\raise.15ex\hbox{/}\mkern-9mu {\bar\partial}}}
\def\pslash{\,\,{\raise.15ex\hbox{/}\mkern-9mu p}}
\def\calDslash{\,\,{\raise.15ex\hbox{/}\mkern-12mu {\cal D}}}

\makeatletter\@addtoreset{equation}{section}\makeatother

\renewcommand{\title}[1]{\vbox{\center\LARGE{#1}}\vspace{5mm}}
\renewcommand{\author}[1]{\vbox{\center#1}\vspace{5mm}}
\newcommand{\address}[1]{\vbox{\center\em#1}}

\def\arXiv#1{\href{http://arxiv.org/abs/#1}{arXiv:#1}}
\def\arXiv#1#2{\href{http://arxiv.org/abs/#1}{arXiv:#1}}

\def\doi#1#2{\href{http://doi.org/#1}{#2}}

\begin{document}

\unitlength = .8mm

\begin{titlepage}
\vspace{.5cm}
 
\begin{center}
\hfill \\
\hfill \\
\vskip 1cm

\title{\boldmath Internal structure of hairy rotating black holes \\in three dimensions
}
\vskip 0.5cm
{Ling-Long Gao$^{\,a,b}$}\footnote{Email: {\tt linglonggao@buaa.edu.cn}},
{Yan Liu$^{\,a,b}$}\footnote{Email: {\tt yanliu@buaa.edu.cn}} and
{Hong-Da Lyu$^{\,a,b}$}\footnote{Email: {\tt hongdalyu@buaa.edu.cn}}

\address{${}^{a}$Center for Gravitational Physics, Department of Space Science\\ and International Research Institute
of Multidisciplinary Science, \\Beihang University, Beijing 100191, China}

\address{${}^{b}$Peng Huanwu Collaborative Center for Research and Education, \\Beihang University, Beijing 100191, China}

\end{center}
\vskip 1.5cm

\abstract{We construct hairy rotating black hole solutions in three dimensional Einstein gravity coupled to a complex scalar field. When we turn on a real and uniform source on the dual CFT, the black hole is stationary with two Killing vectors and we show that there is no inner horizon for the black hole and the system evolves smoothly into a Kasner universe. When we turn on a complex and periodic driving source on the dual CFT with a phase velocity equal to the angular velocity of the black hole, we have a time-dependent black hole with only one Killing vector. We show that inside the black hole, after a rapid collapse of the Einstein-Rosen bridge, oscillations of the scalar field follow. Then the system evolves into the Kasner epoch with possible Kasner inversion, which occurs in most of the parameter regimes. In both cases, one of the metric fields obeys a simple relation between its value at the horizon and in the Kasner epoch. 
}
\vfill

\end{titlepage}

\begingroup 
\hypersetup{linkcolor=black}
\tableofcontents
\endgroup





\section{Introduction}

Recently, there has been much progress in understanding the interior geometries in static black holes and novel behaviors have been found, e.g. \cite{Frenkel:2020ysx, Hartnoll:2020rwq, Hartnoll:2020fhc}.  The purpose of this work is to go beyond static black hole solutions to stationary black hole solutions and time-dependent black holes to study their internal geometries. 

Stationary black hole solutions are more generic than static black hole solutions, i.e. 
static black holes are special limits of 
stationary black holes with time reversal symmetry preserving. Time-dependent black holes are even more generic in which the timelike Killing vector $\partial_t$ is no longer a Killing vector. It is known that with less symmetry or less Killing vector the solution of Einstein's gravity is much richer and more complicated. For example, the stationary Kerr black hole has more complicated exterior and interior structure than the static Schwarzschild black hole. Therefore it is interesting to study the interior behavior of stationary black holes. 

The interior of four and five dimensional rotating black holes with scalar hair in the asymptotic flat case has been studied in \cite{Brihaye:2016vkv} and the nonexistence of the inner horizon has been found. We will be interested in asymptotic AdS black holes. For simplicity, we will consider the three dimensional hairy rotating black holes. It is known that rotating BTZ black hole exists in three dimensions \cite{Banados:1992wn} and it has been served as a candidate of a consistent quantum gravity model \cite{Hawking:1998kw,  Birmingham:2001dt}. Naturally any  deformation of this toy model should be interesting to explore. 

We will consider the three dimensional rotating black hole solution with a complex scalar hair of the form $\phi(z) e^{-i\omega t+i nx }$ in the Einstein-scalar theory.  
Similar profile of the scalar field has been used for constructing the five dimensional rotating black hole with only one Killing vector \cite{Dias:2011at}. In \cite{Stotyn:2012ap, Stotyn:2013spa} attempts to construct three dimensional black holes with only one Killing vector in Einstein gravity coupled to a massless complex scalar field have been made and they concluded that no such perturbative black hole exists. Different from  \cite{Stotyn:2012ap, Stotyn:2013spa} we consider Einstein gravity coupled to a massive complex scalar field and turn on a source on the dual theory to construct the black hole solutions. 

We will focus on the internal structure of the hairy rotating black holes. The internal structure of hairy charged black holes has been studied in \cite{Hartnoll:2020rwq, Hartnoll:2020fhc}. Without any scalar hair, charged black holes and rotating black holes have similar Penrose diagrams. However, they are different in symmetries and the formation of them are completely different too. Thus it deserves to explore the internal structure of rotating black holes with scalar hair.

The paper is organized as follows. In Sec.   \ref{sec:model} we show the detailed setup of the model. In Sec.  \ref{sec:ihr}, we explore the internal structure of hairy rotating blck holes in two different cases. We conclude and discuss the results in Sec.  \ref{sec:cd}.  In Appendix \ref{sec:lt} we show the low temperature solution and in appendix \ref{app:nonih} we provide arguments of no inner horizon for black holes of case II in 
Sec. \ref{subsec:cII}.

\section{Setup of the model}
\label{sec:model}

We consider three dimensional gravity coupled to a complex scalar field with the following action
\begin{align}
    S = \int d^3x \sqrt{-g} \left( R +2 -  \partial_a \varphi \partial^a \varphi^* - m^2 \varphi\varphi^* \right)\,.
\end{align}
 In principle, one could generalize the potential for the complex scalar field $\varphi$ to arbitrary form and here for simplicity we only consider the mass term. 
For convenience we have set $16\pi G=1$ and fixed the cosmological constant. 

The corresponding equations of motion are 
\begin{eqnarray}
\begin{split}
\label{eq:3d}
    R_{ab}-\frac{1}{2} R g_{ab} - g_{ab} -\frac{1}{2}  (\partial_a \varphi \partial_b \varphi^* + \partial_a \varphi^* \partial_b \varphi) + \frac{1}{2}g_{ab} \left(  \partial_c \varphi \partial^c \varphi^* + m^2 \varphi\varphi^* \right) &=0\,, \\
    \nabla_a\nabla^a \varphi - m^2\varphi &=0\,.
\end{split}
\end{eqnarray} 
The ansatz of hairy rotating black hole is given by 
\begin{eqnarray}
\label{eq:ans3d}
\begin{split}
    ds^2 &= \frac{1}{z^2} \left( -f e^{-\chi} dt^2 + \frac{dz^2}{f} + (N dt + dx)^2 \right )\,,\\
    \varphi &= \phi(z) e^{-i \omega t + i n x}\,.
\end{split}
\end{eqnarray}
Here $f, \chi, N, \phi$ are functions of $z$. We assume the spatial direction is periodic, i.e. $x\sim x+2\pi$.\footnote{If we assume $x$ is non-compact, then the solution we considered can be explained as a boosted hairy black hole. } In this case $n$ should be an integer. 
We also assume that $\omega$ is real. 
Note that multiplying the scalar field $\varphi$ by a constant phase does not change the solution.  

For the simplest case $\omega=n=0$, i.e. the scalar field is real, both the time and space translational symmetries are preserved. For nonzero $\omega$ and nonzero $n$, and both the translational symmetry along $t$ and $x$ directions are broken up to a discrete symmetry $t\to t+\frac{2\pi}{\omega}, x\to x+\frac{2\pi}{n}$. In this case the discrete symmetry along $t$ direction reminds us 
the concept of time crystal \cite{Zaletel:2023aej}.  
The solution with $\omega=0, n\neq 0$ is a fine-tunning 
result of this case.  
For the case $n=0, \omega\neq 0$ we have checked that there is no hairy black hole solution while boson star solutions could exist.  
We will focus on the two cases of $\omega=n=0$ (case I) and  $n\neq 0$ (case II) in the following and study the black hole interiors in the next section. 

Note that the ansatz \eqref{eq:ans3d} is invariant under two scaling symmetries
\begin{align}\label{eq:scasym1}
 t\rightarrow \lambda t\,, ~N\rightarrow \lambda^{-1} N\,, ~\chi \rightarrow \chi+2 \log{\lambda}\,, ~\omega\rightarrow \lambda^{-1} \omega\,,
\end{align}
\begin{align}\label{eq:scasym2}
 (t,\,x, \,z) \rightarrow \lambda\,(t,\,x,\,z)\,, ~(\omega,\,n) \rightarrow \lambda^{-1}(\omega,\,n)\,,
\end{align}
and a gauge symmetry
\begin{align}\label{eq:gausym}
    N \rightarrow N+\lambda\,, ~x\rightarrow x- \lambda t\,, ~\omega \rightarrow \omega - \lambda n\,.
\end{align}

Substituting \eqref{eq:ans3d} into \eqref{eq:3d}, we obtain the equations of motion
\begin{align}
\label{eq:eombg}
    \begin{split}
       \phi ''+\phi' \left(\frac{f'}{f}-\frac{\chi '}{2}-\frac{1}{z}\right) + \frac{ e^{\chi} \phi}{f^2}(\omega+nN)^2 - \frac{\phi}{z^2 f}(m^2+n^2z^2) &= 0\,,\\
       \chi '-\frac{2 f'}{f}+\frac{2m^2\phi^2}{z f}-\frac{4}{z f}+\frac{4}{z}+\frac{z e^{\chi } N'^2}{f} + \frac{2z n^2 \phi^2}{f} &=0\,,\\
       \chi '-2z \phi'^2 - \frac{2z e^{\chi} \phi^2}{f^2}(\omega+nN)^2 &=0\,,\\
       N''+ N' \left(\frac{\chi '}{2}-\frac{1}{z}\right) - \frac{2n\phi^2}{f}(\omega+nN) &=0\,,
    \end{split}
\end{align}
where $'$ is the derivative with respect to $z$. 
We have an additional second order ODE for $f$ and $\chi$ which can be derived from above.

The series expansion of the solutions near the (event) horizon $z\to z_h$ are 
\begin{align}
\label{eq:irexp}
    \begin{split}
        f &= \left( \frac{m^2\phi_h^2-2}{z_h} +z_h\, \big(\frac{e^{\chi_h}N_1^2}{2}+n^2 \phi_h^2\big) \right) (z-z_h)+\cdots\,,\\
        \chi &= \chi_h + \frac{8\phi_h^2\left(e^{\chi_h}n^2N_1^2z_h^4+(m^2+n^2z_h^2)^2\right)}{z_h\left(e^{\chi_h}N_1^2z_h^2+2(m^2+n^2z_h^2)\phi_h^2-4\right)^2}\,(z-z_h) +\cdots\,,\\
        N &= N_h + N_1 (z-z_h)+\cdots \,,\\
        \phi &= \phi_h + \frac{2\phi_h\left(m^2+n^2z_h^2\right)}{z_h(e^{\chi_h}N_1^2z_h^2+2(m^2+n^2z_h^2)\phi_h^2-4)} (z-z_h) +  \cdots\,,
    \end{split}
\end{align}
with $N_h=-\frac{\omega}{n}$ for nonzero $n$. 
Therefore for nonzero $n$ 
the angular velocity of the horizon is 
\be\label{eq:syncon}
\Omega=-N_h=\frac{\omega}{n}
\ee
which is exactly the same as the phase velocity of the scalar field. This is similar to the hairy black holes with synchronized hair in asymptotic flat case \cite{Herdeiro:2014goa} and the five dimensional black holes with only one Killing vector  in \cite{Dias:2011at}. 
 When $n=\omega=0$, $N_h$ is a free parameter which can not be fixed. In the case $n=0, \omega\neq 0$, we have $\phi=0$ which indicates that the solution has to be a BTZ black hole. 
 
 In the expansion \eqref{eq:irexp}, we have assumed the existence of the horizon. The system might allow the solutions of boson star, whose geometry does not have horizon. Then in the center of the star we have the boundary condition for generic $n\neq 0, \omega\neq 0$, \footnote{Note that the most general solution for $f\,, \chi\,, \phi$ should be $f= f_0 z^2+ \mathcal{O}(1)\,, \chi = \chi_0 + \mathcal{O}(1/z^2)\,, \phi=\mathcal{O}(1/z^a)$ where $a$ depends on both $f_0$ and $n$. We have chosen $f_0=1\,, \chi_0=0$ to set the spacetime at the center of the star to be flat.}
 \be\label{eq:bndbs}
 f= z^2+ \mathcal{O}(1)\,,~~~ \chi = \mathcal{O}\Big(\frac{1}{z^2}\Big)\,,~~~ N=N_0 + \mathcal{O}\Big(\frac{1}{z^2}\Big)\,, ~~~\phi=\mathcal{O}\Big(\frac{1}{z^n}\Big)\,.
 \ee
 For the case $n=0$,  we found that $N=0$ while the other fields have the same boundary condition as \eqref{eq:bndbs}, which indicates that a non-rotating boson star solution is expected  \cite{Astefanesei:2003rw}. 
The boson star solution for the case with massless complex scalar field has been studied in \cite{Stotyn:2013spa}. The holographic dual of boson stars might be linked to the scar states which is a non-thermalized state in a quantum many-body system \cite{Milekhin:2023was}.
We leave the detailed solutions of the stars and their holographic duals for future study and focus on the black hole solution in the following. 

The asymptotic AdS$_3$ requires that the mass of the scalar field is above the
Breitenlohner-Friedman (BF) bound, i.e. $m^2>-1$. We choose $m^2=-3/4$ from here on. Near the boundary $z\to 0$, the metric fields and scalar field are   
\bea
\label{eq:bndexp}
\begin{split}
    f &= 1+ \frac{\phi_0^2}{2} z + f_2 z^2 + \frac{3\phi_0^4}{8} z^2 \log z+\cdots\,,\\
    \chi &= \chi_b + \frac{\phi_0^2}{2} z + \frac{1}{16} (\phi_0^4+24\phi_0 O) z^2 +\frac{3\phi_0^4}{8} z^2 \log z+\cdots\,,\\
    N &= N_b + N_2 z^2+\cdots\,,\\
    \phi &= \sqrt{z}\, \Big( \phi_0 + O z + \frac{\phi_0^3}{4} z \log z - \frac{\phi_0}{128} \left(16 f_2 - 64n^2 + 21\phi_0^4 -48\phi_0 O +64e^{\chi_b} (\omega+n N_b)^2\right)z^2\\ 
    & ~~~~~~+ \frac{3\phi_0^5}{64} z^2 \log z+\cdots  \Big)\,.
\end{split}
\eea

The above UV expansion can be easily generalized to any  
relevant deformation of scalar field, i.e. $m^2<0$. 
When $m^2=0$, for $n=\omega=0$, i.e. the real scalar, we find that the solution should be a rotating BTZ black hole with a constant scalar field; while for 
$n\neq 0$, the hairy black hole solution exists.\footnote{Note that the black hole solution exists only when we turn on a source for the scalar field. This is consistent with the previous result of \cite{Stotyn:2012ap} that black hole solution does not exist for sourceless boundary condition of the scalar field when $m^2=0$.}  
When $m^2>0$, the source term of the scalar field 
will strongly backreact to the metric field near the boundary, 
the leading geometry is no longer AdS$_3$ 
at $z=0$ unless the scalar field is sourceless, i.e. $\phi_0=0$. We have not found such kind of solution.

We can solve the system using shooting method. Note that the IR expansion \eqref{eq:irexp} for the case of nonzero $n$ has six free parameters\footnote{Here we treat $n$ and $\omega$ as free parameters since the scaling symmetries of the system involve the transformation of them.} $z_h, \chi_h, \phi_h, \omega, n, N_1$ but only three of them are independent due to the symmetries \eqref{eq:scasym1}, \eqref{eq:scasym2}, and \eqref{eq:gausym}. This is consistent with the fact that there are three independent scale invariant $T/J, \phi_0/\sqrt J, \omega/n$ in the boundary field theory. Numerically, we use symmetry \eqref{eq:scasym1}, \eqref{eq:scasym2}, and \eqref{eq:gausym} to fix $\chi_h=0, n=1$ and $\omega=1$ respectively. Then we choose the initial value for our shooting parameter $z_h, N_1, \phi_h$ to obtain the black hole exterior and interior solutions, from which we could further rescale to the solution we will work in as follows. Here we work in the Schwarzschild coordinates, i.e. the frame that does not rotate at the boundary. The dual field theory is defined on the asymptotic AdS$_3$ boundary with $ds^2=-dt^2+dx^2$.\footnote{In what follows, we work in this specific coordinate.} This can be achieved from the above black hole solution by using the symmetry \eqref{eq:gausym} to fix $N_b=0$ and using \eqref{eq:scasym1} to set $\chi_b=0$.

The dual field theory is a two dimensional CFT defined in a cylinder at finite temperature with a nontrivial angular momentum and a deformation with a scalar source of form $\phi_0 e^{-i\omega t+i n x}$ turning on. 
We have three dimensionless quantities $T/J, \phi_0/\sqrt{J}, \omega/n$\footnote{Here the parameters $\omega, n$ could not be viewed as thermodynamical quantities, while they play important roles in energy and momentum relaxation.} 
where $J, T, \phi_0$
 are the angular momentum, the temperature and the scalar source respectively. The temperature is 
\be
T=-\frac{f'e^{-\chi/2}}{4\pi}\bigg|_{z=z_h}\,.~~\\
\ee
The source $\phi_0$ can be obtained from the boundary expansion in  \eqref{eq:bndexp}. The mass and angular momentum of the hairy black hole can be calculated to be 
\be
\label{eq:mj}
\begin{split}
M&=-\lim_{z\rightarrow 0} \sqrt{-h}\, T^t_{~t}=\frac{1}{4}e^{-\chi_b/2}(-4f_2 + 8 e^{\chi_b} N_2 N_b + \phi_0^4 + 6\phi_0 O)\,,~~~\\
J&=\lim_{z\rightarrow 0} \sqrt{-h}\, T^t_{~x}=-2\,e^{\chi_b/2}N_2\,,~~~
\end{split}
\ee
where $h$ is the determinant of the induced metric on the boundary and $T_{ab}$ is the energy momentum tensor for the dual field theory. It takes the form of   
\begin{align}
T_{ab} = 2(K_{ab} -K h_{ab} -h_{ab}) +\frac{1}{2}h_{ab}\,(\varphi^* n^c \partial_c \varphi -\frac{1}{2}\varphi^* \varphi)   \,, 
\end{align}
obtaining from holography using the counter-term \cite{Gegenberg:2003jr} 
\begin{align}
S_\text{ct}= - \int_{\partial\mathcal{M}} d^2x \sqrt{-h} \left(2 K +2 - \frac{1}{2} \varphi^* n^c \partial_c \varphi + \frac{1}{4} \varphi^* \varphi \right)
\end{align}
where $n^a$ is the normal vector of boundary $\partial\mathcal{M}$ which is located at  $z=0$, $K_{ab}=-(\nabla_a n_b+ \nabla_b n_a)/2$ is the extrinsic curvature tensor and $K=h^{ab}K_{ab}$ is the extrinsic curvature scalar.


The system allows a rotating BTZ solution with a trivial scalar field \cite{Banados:1992wn}, 
\begin{align}
\label{eq:btz}
    f= \frac{(z^2-z_h^2)(z^2-z_i^2)}{z_h^2z_i^2}\,, 
    ~~~\chi= 0\,,
    ~~~N= -\frac{z^2}{z_hz_i}\,, 
    ~~~\phi=0 \,.
\end{align}
The angular velocity of BTZ black hole is $\Omega=z_h/z_i$, stability requires $\Omega\leq 1$. (i.e. $M=\frac{1}{z_h^2}+\frac{1}{z_i^2},$ $J=\frac{2}{z_h z_i}$, with $M\geq J$.) 
From the analysis of the quasi-normal modes (QNM) of the scalar field around rotating BTZ black holes, it is known that the system is stable for $m^2>-1$ \cite{Birmingham:2001hc}. However,  
for rotation black holes, there might be superradiant instability for the free complex scalar field around rotating BTZ black holes when $0<\omega< -n N_h$ 
if we impose the mixed (or Robin) boundary conditions while not Dirichlet or Neumann boundary condition for the complex scalar field 
\cite{Ortiz:2011wd, Dappiaggi:2017pbe} or real scalar field \cite{Iizuka:2015vsa}. 
Interestingly clouds with nontrivial configurations of scalar fields exist at the threshold of superradiance \cite{Dappiaggi:2017pbe}. 
Moreover, it was shown in \cite{Balasubramanian:2004zu, Dias:2019ery, Emparan:2020rnp, Kolanowski:2023hvh} that the scalar perturbations around non-extremal rotating BTZ should lead to an instability of the inner horizon. From these results obtained in the probe limit, we expect that it is important to consider the backreaction of the scalar field to the rotating BTZ black hole. The hairy rotating black hole solutions we will construct in the next section  could be viewed as a possible endpoint of the above instabilities in the probe limit.

\section{Inside the hairy rotating black holes}
\label{sec:ihr}

In this section, we will consider the black hole solutions in two different cases. In the first case of $\omega=n=0$, the black hole solution is stationary with two isometries. The time reversal symmetry is broken compared to the static BTZ black hole. The second case is for $\omega\neq 0, n\neq 0$ where the time-dependent black hole has only one Killing vector. We will study the internal structures of these black holes. 

\subsection{Case I: when $\omega=n=0$}
\label{subsec:cI}

In this case, the scalar field of the hairy rotating black hole is real. In the dual field a real and uniform source is turned on. In the following we will first collect the useful equations and then study the internal structure both numerically and analytically. 

The ansatz \eqref{eq:ans3d} becomes 
\bea
\label{eq:ans3d-c1}
\begin{split}
    ds^2 &= \frac{1}{z^2} \left( -f e^{-\chi} dt^2 + \frac{dz^2}{f} + (N dt + dx)^2 \right )\,,\\
    \varphi &= \phi(z) \,,
\end{split}
\eea
which satisfies the following equations
\begin{align}
\label{eq:eom3d-c1}
    \begin{split}
       \phi ''+\phi' \left(\frac{f'}{f}-\frac{\chi '}{2}-\frac{1}{z}\right)-\frac{m^2\phi}{z^2 f}  &= 0\,,\\
       \chi '-2 z \phi'^2 &=0\,,\\
       \chi '-\frac{2 f'}{f}+\frac{2m^2\phi^2}{z f} -\frac{4}{z f}+\frac{4}{z}+\frac{z e^{\chi } N'^2}{f} &=0\,,\\
       N''+ N' \left(\frac{\chi '}{2}-\frac{1}{z}\right) &=0\,.
    \end{split}
\end{align}

There are three symmetries
for the system
\begin{align}
 t\rightarrow bt\,, ~N\rightarrow N/b\,, ~\chi \rightarrow \chi+2 \log{b}\,;\\
 (t,\,x, \,z) \rightarrow b\,(t,\,x,\,z)\,;\\
 N\to N+b\,,~ x\to x-bt\,.
\end{align}
We can use the first and third symmetry to set the leading order of $N$ and $\chi$ to be zero at boundary $z=0$. We can also set $z_h=1$ from the second symmetry above when the location of the horizon $z_h$ is defined from $f(z_h)=0$. The corresponding conserved charge from the above symmetries is 
$
Q = e^{\chi/2}\left(N^2 - f e^{-\chi}\right)'/z\,,
$ which can be used to verify the numerical accuracy.

There are two independent Killing vectors $\partial_t$ and $\partial_x$ in this stationary case. The combination $\xi=\partial_t-N_h \partial_x$ gives null Killing vectors on the horizon where $N_h$ is the value of $N$ at the horizon $z_h$.  
Note that different from the case $n\neq 0$, there is no constraint on the value of $N_h$ in this case. 

\begin{figure}[h!]
\begin{center}
\includegraphics[width=0.41\textwidth]{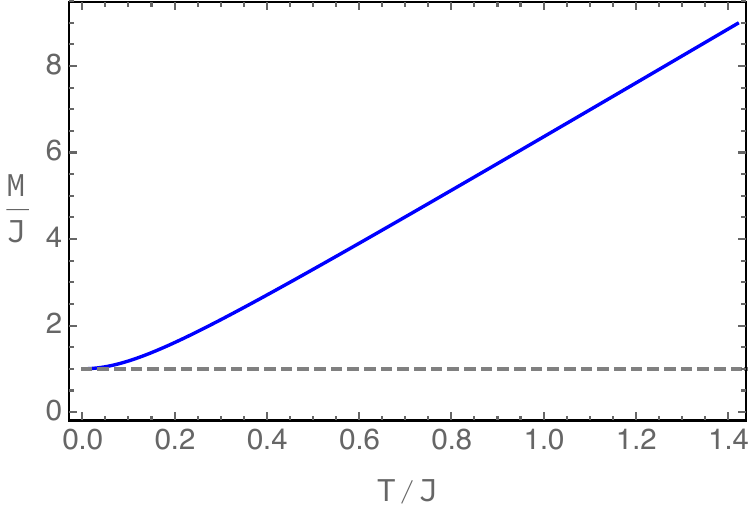}
~~~
\includegraphics[width=0.426\textwidth]{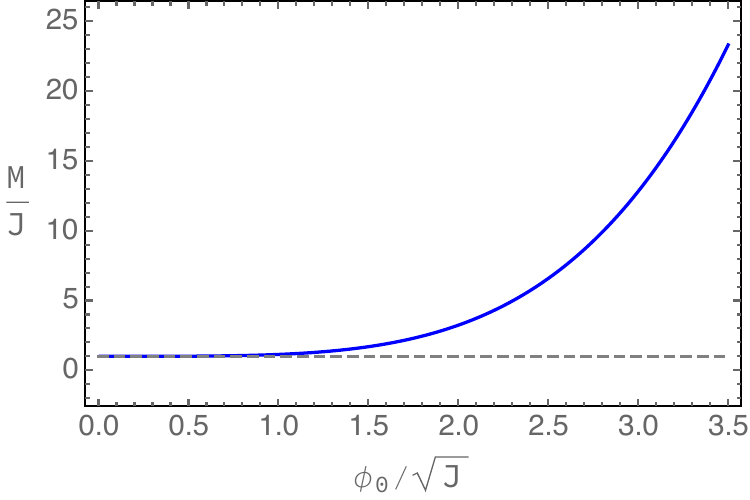}
\end{center}
\vspace{-0.3cm}
\caption{\small Plot of $M/J$ as a function of $T/J$ at $\phi_0/\sqrt J=0.05$ ({\em left}) and as a function of $\phi_0/\sqrt J$ at $T/J=0.01$ ({\em right}). }
\label{fig:mbjneu}
\end{figure}

The near horizon condition and the near boundary condition are the same as \eqref{eq:irexp} and (\ref{eq:bndexp}), except that $N_h$ is a free parameter now. Using the above symmetries, there are two free parameters near the horizon, which correspond to $T/J$ and $\phi_0/\sqrt{J}$ in the dual field theory. One can integrate the system into the boundary to obtain the numerical solutions for the black holes. In Fig. \ref{fig:mbjneu}, we show 
$M/J$ as functions of $T/J$ for fixed $\phi_0/\sqrt{J}$ and of  $\phi_0/\sqrt{J}$ for fixed $T/J$ respectively. 
It is interesting to note that all the solutions we studied satisfy the relation $M/J\geq 1$. In the extremal limit, we have $M/J\to 1$. Interesting behavior for the profiles of the metric fields is found at extremely low temperature, as shown in the appendix \ref{sec:lt}.

 We find that the metric field $N$ is a monotonic function from the interior to the boundary numerically. This can also be seen analytically as follows.  
 From the fourth equation in \eqref{eq:eom3d-c1} 
 we have 
 $N'=E_0 z e^{-\chi/2}$,
 where $E_0$ is an integration constant. Thus the function $N(z)$ is monotonically increasing or decreasing. 
 From \eqref{eq:mj} we have $J=-E_0$. We consider the case with positive $J$, which means that $N$ is monotonically decreasing from the interior to the boundary.

\subsubsection{No inner horizon 
}\label{sec:noinneu}

Without constructing a detailed numerical solution, we can prove that there is no inner horizon for the hairy black hole solution for $m^2\leq 0$. 
The first equation of \eqref{eq:eom3d-c1} can be written as
\begin{align}\label{eq:eomphi3d}
   \left(\frac{f e^{-\chi/2} \phi'}{z}\right)' &= \frac{e^{-\chi/2}}{z^3} m^2\phi\,.
\end{align}
If there were two horizons $f(z_h) = f(z_i) = 0 $ with outer horizon $z_h$ and inner horizon $z_i$, 
from \eqref{eq:eomphi3d} we would have
\begin{align}
    0= \int_{z_h}^{z_i} \left( \frac{fe^{-\chi/2}\phi\phi'}{z} \right)' dz = \int_{z_h}^{z_i} \frac{e^{-\chi/2}}{z^3} \left( z^2 f \phi'^2+ m^2\phi^2 \right) dz\,.
\end{align}
The first equality is because of  $f(z_h) = f(z_i) = 0 $. However, since we have $f(z) < 0$
between the two horizons and $m^2 < 0$ which implies the integrand of right hand side is negative.
Therefore there can not be an inner horizon for $m^2<0$. 

When $m^2=0$, we only have the trivial solution of BTZ black hole with a constant scalar field. This can be seen as follows. The equation of scalar field $\phi$ can be solved as 
\begin{align}
    \phi'= C\, z f^{-1} e^{\chi/2}
\end{align}
where $C$ is a constant. Then the third equation in  \eqref{eq:eom3d-c1} becomes
\begin{align}
   2 C^2 z^3 e^{\chi} = f^2 \chi'\,.
\end{align}
Because $f=0$ at the horizon $z=z_h$, we must have $C=0$ which means that the scalar field is constant. 
Therefore we only have a BTZ black hole solution \eqref{eq:btz} 
with a constant scalar field 
when $m^2=0$.

When $m^2>0$, the deformation of the scalar operator becomes relevant and the asymptotic AdS structure is destroyed unless the source $\phi_0=0$. Numerically we have not found any sourceless solution for this case and this is consistent with the fact that there is no instability in the bulk.

\subsubsection{Collapse of the Einstein-Rosen bridge}
We have proved that there is no inner horizon for $m^2\leq 0$ in the above subsection. Numerically we found that the inner horizon $z_I$ of the rotating BTZ black hole at the same value of $T/J$ collapses. The collapse can be approximately described analytically near $z_I$ as follows. Note that the analytical calculation in the following subsections works for any value of $m^2$ while the numerical results are for $m^2=-3/4$. 

The first three equations in \eqref{eq:eom3d-c1} can be rewritten as
\begin{align}
\label{eq:eom3d-c1re}
\begin{split}
    z^3 e^{\chi/2} \left(\frac{e^{-\chi/2} f  \phi'}{z}\right)' &= m^2\phi\,,\\
    \chi' &= 2 z(\phi')^2\,,\\
    2z^3 e^{\chi/2} \left(\frac{e^{-\chi/2}f}{z^2}\right)' &= -4+2m^2\phi^2 + z^4 E_0^2\,,
\end{split}
\end{align}
where we have used the fourth equation of \eqref{eq:eom3d-c1} which can be integrated to be
$N'=E_0 z\, e^{-\chi/2}\,, $
with $E_0$ a constant of integration. 
Since the collapse occurs over an extremely short range in the $z$ coordinate,  we can set $z=z_*$ to be the would-be inner horizon where $z_*$ is a position very close to the inner horizon $z=z_I$ of corresponding BTZ black hole at same value of $T/J$. We can also think $f, \chi, \phi$ as functions of $\delta z = z-z_*$. Furthermore, we have numerically checked that the mass term of scalar field can be neglected due to the small value of $\phi$ in the first and third equations in \eqref{eq:eom3d-c1re} comparing other terms\footnote{\label{ft:neglectscalar}Numerically, we have verified that the mass term of the scalar field $\phi$ is on the  order of $10^{-3}$ compared to the other terms near the would-be inner horizon, allowing it to be safely  neglected.}. Then the equations \eqref{eq:eom3d-c1re} become   
\begin{align} 
    \label{cerb-simplified}
    \begin{split}
    \left(e^{-\chi/2} f  \phi'\right)' &= 0\,,\\
    \chi' &= 2 z_*(\phi')^2\,,\\
    (e^{-\chi/2} f)' &= \frac{E_0^2}{2} z_*^3 e^{-\chi/2} - 2\frac{e^{-\chi/2}}{z_*}\,.
     \end{split}
\end{align}

We can integrate the first equation in \eqref{cerb-simplified} to obtain 
$\phi '= - a \,e^{\chi/2}/f$ where $a$ is a constant of  integration. Plugging it back into \eqref{cerb-simplified} and taking the derivative of the last equation 
we obtain 
\begin{equation}
    \label{eq-gtt}
    \frac{g_{tt0}''}{g_{tt0}'}=\frac{c_1^2 g_{tt0}'}{g_{tt0}\left(c_1^2+g_{tt0}\right)}\,,
\end{equation}
where $g_{tt0}=-f e^{-\chi}/z_*^2$ and
$c_1=\sqrt{2a^2/(z_*^4 E_0^2-4)}\,.$ 
The solution of \eqref{eq-gtt} takes the form
\begin{equation}
    \label{eq:sol-gtt0-neu}
    c_1^2 \log\left(g_{tt0}\right)+g_{tt0}=-c_2^2 \left(\delta z+ c_3 \right),
\end{equation}
where $c_2$ and $c_3$ are constants of integration.\footnote{We use this equation to fit the numerical result of $g_{tt0}$ for a range of $z$.  The parameter $z_*$ is chosen such that $c_3=0$. Equivalently, the fitting parameters are $c_1, c_2, z_*$ while $c_3=0$.  } 
Taking the derivative of \eqref{eq:sol-gtt0-neu} we 
obtain 
$    g_{tt0}'=-c_2^2 g_{tt0}/(c_1^2+g_{tt0})
$
and together with $f= -z_*^2 e^{\chi} g_{tt0}$ we can get the solutions of $\phi\,,\chi$ and $N$
\begin{align}
    \begin{split}
    \phi &= -\frac{c_1}{z_* c_2}\log\left(c_4\, g_{tt0}\right)\,,\\
    e^{-\chi} &= \frac{z_*^{4}}{ a^2}\,\phi'^2 g_{tt0}^2\,,\\
    N' &= \frac{E_0 z_*^3}{a}\left| \phi' g_{tt0}\right|.
    \end{split}
\end{align}
where $c_4$ is a constant of integration.

Collect all the results above we find a fast crossover of the fields  
near $z_*$\footnote{Note the below equations are valid for $0\ll \delta z\ll z_*$.} 
\begin{align}
\label{eq:crossover}
    \begin{split}
    \delta z <0 ~~~& \rightarrow~~~ \delta z >0\,,\\
        g_{tt0}=c_2^2\left|\delta z\right| ~~~& \rightarrow~~~ g_{tt0}=e^{-\left( c_2^2 /c_1^2\right)\delta z}\,,\\
        \phi'= \frac{-c_1}{z_* c_2} \frac{1}{\left|\delta z\right|} ~~~& \rightarrow ~~~\phi'=\frac{c_2}{c_1}\,,\\
        N'=\frac{E_0 c_2 z_*^3 }{a} ~~~& \rightarrow~~~
        N'=\frac{E_0 c_2 z_*^3 }{c_1^2 a} e^{-\left( c_2^2/c_1^2\right)\delta z}\,.
    \end{split}
\end{align}
The $g_{tt0}$ is linearly vanishing for $\delta z<0$ which is similar to the behavior of the rotating BTZ black hole when it goes to the inner horizon. When $\delta z>0$, $g_{tt0}$ decays exponentially to a nonzero value in extremely small range about $\Delta z= (c_1/c_2)^2$.
Note that although $g_{tt0}$ is only one term in the metric component $g_{tt}= (-f e^{-\chi}+N^2)/z^2$, but the volume element of the wormhole connecting the two exteriors of the black hole depends on the determinant of induced metric on it which is precisely $\sqrt{g_{tt0}}$. 
Therefore it could be used to measure the size of the Einstein-Rosen bridge, 
which is similar to the collapse of the Einstein-Rosen bridge for the static black holes in \cite{Hartnoll:2020fhc, Hartnoll:2020rwq}.  

\begin{figure}[h!]
\begin{center}
\includegraphics[width=0.50\textwidth]{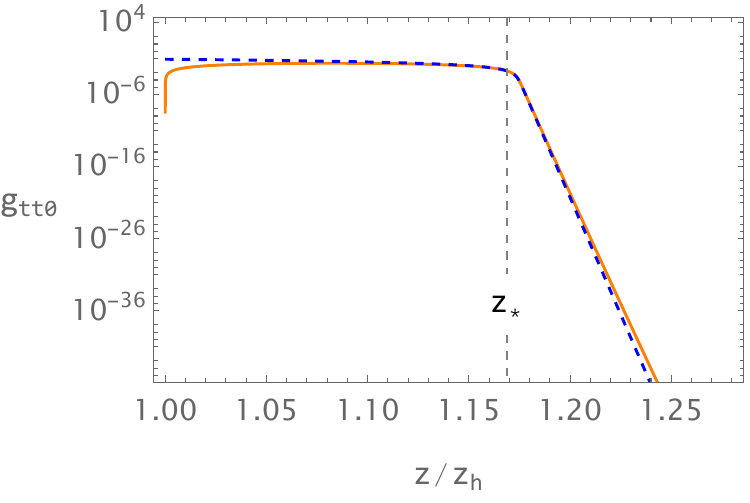}
\end{center}
\vspace{-0.4cm}
\caption{\small  Illustration of Einstein-Rosen bridge collapse. The solid orange line is the numerical solution of \eqref{eq:eom3d-c1} and the dashed blue line is the solution of \eqref{eq:sol-gtt0-neu}. In this case, we have $z_*=1.17\,,\  c_2/c_1 = 50.58$ and the parameters of the solution are $T/J=0.025\,,\  \phi_0/\sqrt{J}=0.071$. }
\label{fig:ercneu}
\end{figure}

We illustrate the 
Einstein-Rosen bridge collapse in Fig. \ref{fig:ercneu} which is the $\log$ plot of $g_{tt0}$ as a function of $z/z_h$. It can be seen that near the would-be inner horizon $z=z_*$ the full numerical solution of \eqref{eq:eom3d-c1} and the solution of \eqref{eq:sol-gtt0-neu} match very well. When $z>z_*$, the linear decrease in the plot indicates exponential decay of $g_{tt0}$.

Moreover, from \eqref{eq:crossover} we find that there is no oscillation for the scalar field. This is quite similar to the hairy charged black with a charged neutral scalar field \cite{Hartnoll:2020rwq}.

\subsubsection{Kasner exponents}

After the collapse of the Einstein-Rosen bridge, the system evolves shortly into the Kasner regime where the geometry is characterized by a Kasner universe. 
Different from the BTZ black hole where the center of the black hole is a conical singularity with finite curvature, in the hairy BTZ black hole there is a curvature singularity near the center. 

Near the singularity $z\rightarrow \infty$,  the equations of motion \eqref{eq:eom3d-c1} can be simplified under the assumption that the ignored terms are subleading which will be checked numerically  afterward,
\begin{align}
    \label{eq:simplified eqs}
    \begin{split}
       \phi ''+\phi' \left(\frac{f'}{f}-\frac{\chi '}{2}-\frac{1}{z}\right)  &= 0\,,\\
       N''+ N' \left(\frac{\chi '}{2}-\frac{1}{z}\right) &=0\,,\\
       \chi '-2 z \phi '^2 &=0\,,\\
       \chi '-\frac{2 f'}{f} +\frac{4}{z}&=0\,.
    \end{split}
\end{align}
Eliminating $\chi'$ from the second equation then substituting the result into the first equation we obtain 
\begin{align}
\label{eq:sinsolphineu}
    \phi \sim \alpha \log z
\end{align}
where $\alpha$ is a constant of integration. It has to be determined from the UV data and can only be obtained numerically. 
Plugging \eqref{eq:sinsolphineu} back into \eqref{eq:simplified eqs} we obtain the solution
\begin{align}\label{eq:sinsolneu}
    f \sim -f_K z^{\alpha^2+2}\,, 
    ~~~
    \chi \sim 2\alpha^2 \log z+ \chi_1\,, 
    ~~~
    N \sim N_K + \frac{E_K}{2-\alpha^2}z^{2-\alpha^2}\,,
\end{align}
where $f_K, \chi_1, N_K, E_K$ are constants. Note that in equations \eqref{eq:simplified eqs}, we have assumed that the terms ignored are subleading. More explicitly, we have assumed
\begin{align}
\label{eq:alpcond}
    |\alpha|>\sqrt{2}\,.
\end{align}

Substituting the solutions \eqref{eq:sinsolphineu} and  \eqref{eq:sinsolneu} into ansatz \eqref{eq:ans3d-c1} and performing the coordinate transformation
\begin{align}
    \tau = \frac{2}{\sqrt{f_K} (\alpha^2+2)} z^{-(\alpha^2+2)/2}\,,
\end{align}
we obtain the Kasner form for the fields
\begin{align}
\label{eq:kasnerform1}
\begin{split}
    ds^2 &= -d\tau^2 + c_t \tau^{2p_t} dt^2 + c_x \tau^{2p_x} (N_Kdt+dx)^2\,,\\
    \phi &= \, p_{\phi}\log\tau + c_{\phi}\,,
\end{split}
\end{align}
where the Kasner exponents $p_t, p_x, p_{\phi}$ are
\begin{align}\label{eq:nkas}
   p_t=\frac{\alpha^2}{\alpha^2+2}\,, ~~~~~p_x=\frac{2}{\alpha^2+2}\,, 
    ~~~~~p_{\phi}=-\frac{2 \alpha}{\alpha^2+2}\,, 
\end{align}
which satisfy the Kasner relations
\begin{align}
    p_t+p_x=1\,, 
    ~~~~~p_t^2+p_x^2+p_{\phi}^2=1\,,
\end{align}
and 
\begin{align}
    c_t= f_K e^{-\chi_1}\left( \frac{\sqrt{f_K}(\alpha^2+2)}{2} \right)^{2p_t}\,, ~c_x=\left( \frac{\sqrt{f_K}(\alpha^2+2)}{2} \right)^{2p_x}\,,~ c_{\phi}= p_{\phi} \log \left( \frac{\sqrt{f_K}(\alpha^2+2)}{2} \right)\,.
\end{align}

Note that in \eqref{eq:kasnerform1} we have assumed  that the leading order of $N$
is a constant $N_K$, i.e. $\alpha^2>2$, so that we are able to introduce a new independent coordinate 
$d\tilde{\Omega}=N_K dt+dx$. For the case $0<\alpha^2<2$,  
the Kasner exponents will change under some complicated physical process such as Kasner transition or Kasner inversion. But we have not found this case in the numerical solutions. From \eqref{eq:alpcond} and \eqref{eq:nkas}, we have
\begin{align}
   \frac{1}{2}<p_t<1\,,~~~~~ 
   0<p_x<\frac{1}{2}\,, ~~~~~
   |p_{\phi}|<\frac{2}{3}\,.
\end{align}
Only under this condition can we regard the form \eqref{eq:kasnerform1} as the Kasner form. 

In addition to the fact that the metric function $N$ is a monotonic function from the horizon to the interior, 
from numerics one interesting observation is that we always have $N_K=1/N_h$ where $N_K$ the leading order of $N$ at the Kasner singularity while $N_h$ is the value of $N$ at the horizon. It would be interesting to prove it analytically. 

\begin{figure}[h!]
\begin{center}
\includegraphics[width=0.312\textwidth]{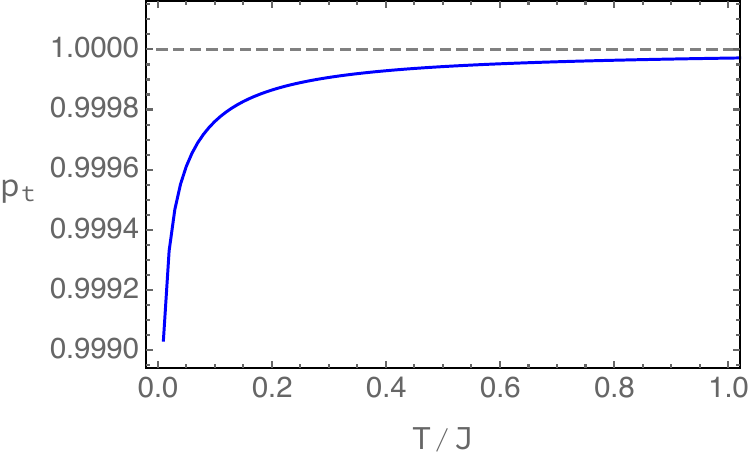}
~
\includegraphics[width=0.322\textwidth]{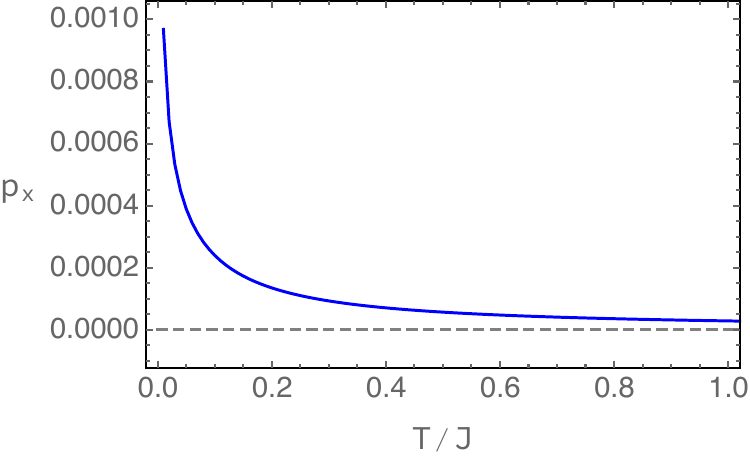}
~
\includegraphics[width=0.312\textwidth]{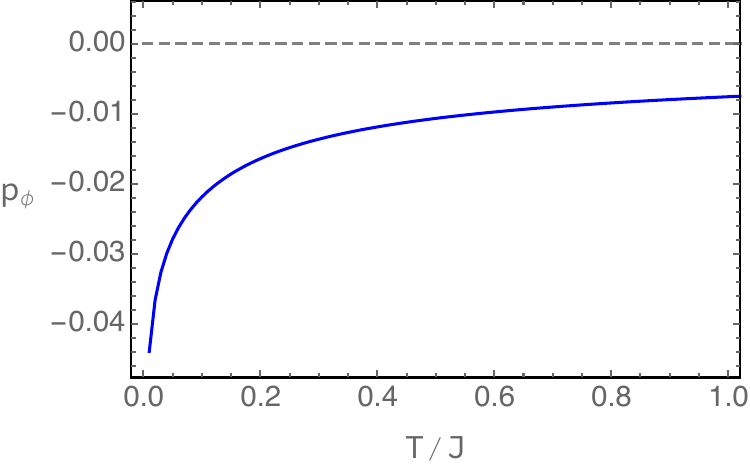}
\end{center}
\vspace{-0.3cm}
\caption{\small The Kasner exponents as functions of $T/J$ for fixed $\phi_0/\sqrt J=0.05$. The dashed gray lines are Kasner exponents of regular inner horizon of BTZ black hole.   }
\label{fig:ptneu}
\end{figure}

In Fig. \ref{fig:ptneu}, we show the dependence of Kasner exponents as functions of $T/J$ for fixed $\phi_0/\sqrt J=0.05$. We find that when $T/J\to \infty$, the  Kanser exponents approach the ones of nonrotating BTZ black hole, i.e. the conical singularity. 

\begin{figure}[h!]
\begin{center}
\includegraphics[width=0.31\textwidth]{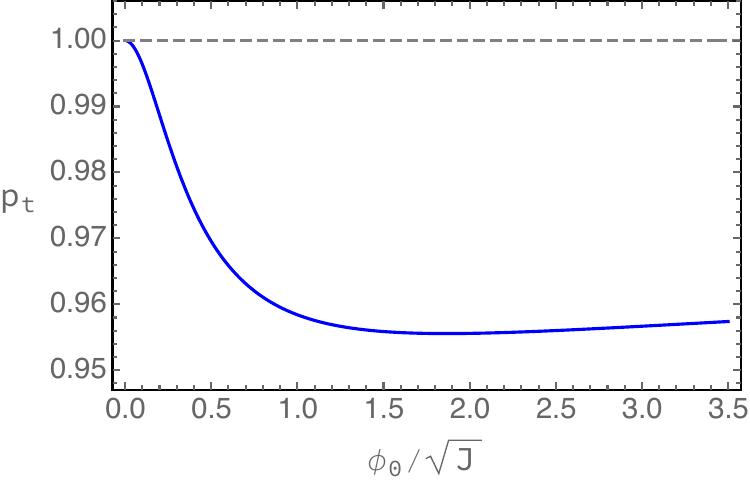}
~
\includegraphics[width=0.315\textwidth]{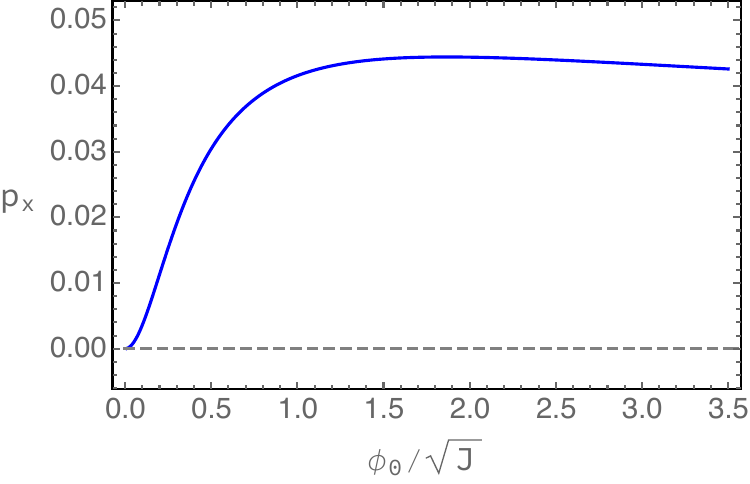}
~
\includegraphics[width=0.323\textwidth]{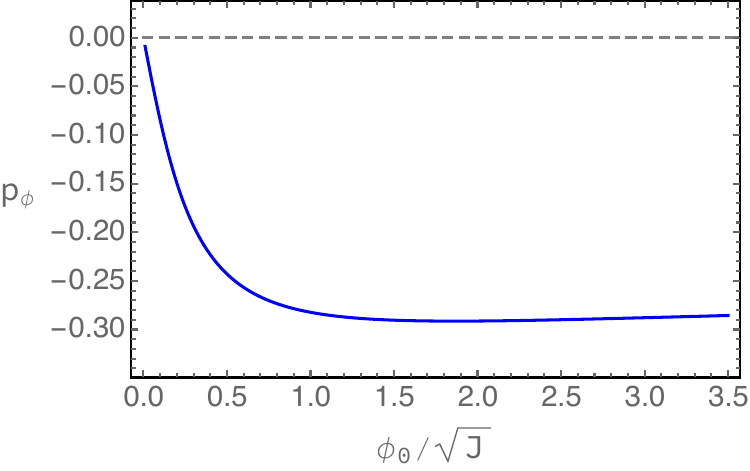}
\end{center}
\vspace{-0.3cm}
\caption{\small The Kasner exponents as functions of $\phi_0/\sqrt J$ for fixed $T/J=0.01$. The dashed gray lines are Kasner exponents of regular inner horizon of BTZ black hole.   }
\label{fig:ptneu1}
\end{figure}

In Fig. \ref{fig:ptneu1}, we show the dependence of Kasner exponents as functions of $\phi_0/\sqrt J$ for fixed $T/J=0.01$. When $\phi_0/\sqrt J \rightarrow 0$, the values of Kasner exponents back to the values of the regular inner horizon of rotating BTZ black hole.\footnote{It is interesting to note that these Kasner exponents are the same as the one in the center of non-rotating BTZ black hole. } This is consistent with the fact that the black hole solution becomes rotating BTZ black hole when $\phi_0/\sqrt J\to 0$.

\subsection{Case II: for general nonzero $n$}
\label{subsec:cII}

In the case of $n\neq 0$, the scalar field in the hairy black hole is complex. In the dual field theory we have turned on a both spatially and time periodic source with a phase velocity equal to the angular velocity of the black hole. In the following again we first list the  useful equations and then study the internal structure of the black holes. 

The ansatz of the background is \eqref{eq:ans3d}. 
The equations of motion can be rewritten as 
\begin{align}
\label{eq:eom3d2}
    \begin{split}
       z^3 e^{\chi/2}\left(\frac{f e^{-\chi/2} \phi'}{z}\right)' &= \phi \left(m^2+n^2z^2-\frac{e^{\chi}}{f}z^2(\omega+n N)^2\right)\,,\\
       z e^{-\chi/2} \left(\frac{e^{\chi/2} N'}{z}\right)' &= \frac{2n\phi^2}{f}(\omega+n N)\,,\\
       \frac{\chi'}{2z} &= \phi'^2 + \frac{e^{\chi}\phi^2}{f^2}(\omega+n N)^2\,,\\
       2e^{\chi/2}z^3 \left(\frac{e^{-\chi/2} f}{z^2}\right)' &= 2(m^2+n^2z^2)\phi^2 + z^2 e^{\chi} N'^2-4\,.
    \end{split}
\end{align}

The Killing vector of this solution can be analyzed as follows. From the metric we have Killing vectors  $\partial_t$ and $\partial_x$. From the scalar field we have $\xi^b\nabla_b\varphi=0$, 
the Killing vector has to be $\xi=\partial_t-N_h \partial_x$ 
with $N_h=-\frac{\omega}{n}$. 
Thus the ansatz \eqref{eq:ans3d} describes a black hole with only one Killing vector, 
\be
\xi=\partial_t+\frac{\omega}{n}\, \partial_x 
\,.
\ee
The norm of the Killing vector is 
\be
g_{ab} \xi^a \xi^b=\frac{1}{z^2}\left(N+\frac{\omega}{n}\right)^2-\frac{f}{z^2}e^{-\chi}\,.
\ee
When $z=z_h$, we have  $g_{ab} \xi^a \xi^b=0$. Therefore the event horizon is also a Killing horizon associated with the Killing vector $\xi$. Just outside the horizon, the Killing vector is always timelike which is different from four dimensional Kerr black hole.  
Near the boundary the Killing vector $\xi^a$ has norm $\frac{1}{z^2}\,(\frac{\omega^2}{n^2}-1)$ which is  spacelike, null or timelike for $\big{|}\frac{\omega}{n}\big{|}>1, =1, <1$ respectively. All the solutions we find are for $\big{|}\frac{\omega}{n}\big{|}<1$ 
which means that the Killing vector is timelike. This behaves differently with the five dimensional rotating black hole with only one Killing vector \cite{Dias:2011at}.

\begin{figure}[h!]
\begin{center}
\includegraphics[width=0.50\textwidth]{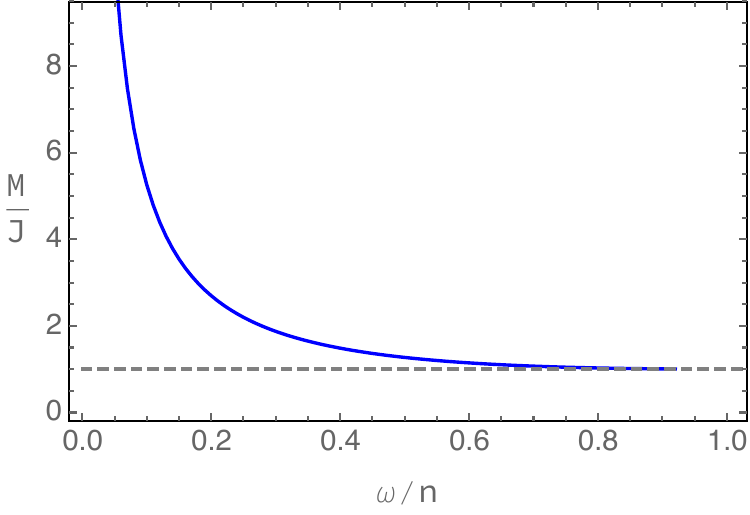}
\end{center}
\vspace{-0.3cm}
\caption{\small Plot of $M/J$ as a function of $\omega/n$ at $\phi_0/\sqrt J=0.5$ and $T/J=0.05$. }
\label{fig:mbjomen}
\end{figure}

In Fig. \ref{fig:mbjomen}, we show $M/J$ as a  function of $\omega/n$ when other parameters are fixed. We find $M/J$ is a monotonic decreasing function of $\omega/n$. When $\omega/n \to 0$ we have $M/J \to \infty$ which can be understood as a non-rotating black hole, because when $\omega=0, n\neq 0$, i.e. $\Omega_h=0$, we only have a non-rotating black hole with $N=0$. When $\omega/n\to 1$, we have $M/J\to 1$. Note that similar to the black holes in section \ref{subsec:cI}, all the black hole solutions we found satisfy 
$M/J\geq 1$. At low temperature, we find the similar behaviors of the system as case I discussed in Sec. \ref{subsec:cI}, which are summarized in Appendix \ref{sec:lt}, i.e. an extremal BTZ black hole emergies with clouds of the scalar field.

Numerically we find that for the solutions we have studied, $N$ is monotonically increasing from the boundary to the interior before the location where Kasner inversion occurs. This observation is not straightforward to proved analytically. Furthermore, we numerically find that the scalar field $\phi$ does not have any node outside the horizon.

Before proceeding, we first have a comment on the non-existence of inner horizon. 
Notice that near the horizon we have $f\to 0$ and $\omega+n N\to 0$. One might attempt to prove the no inner horizon using the methods with equations of motion, or conserved charges. In Appendix \ref{app:nonih} we summarize the attempts 
and show that they do not work here. Nevertheless, one can work in the probe limit and consider a scalar field propagating around the rotating BTZ black hole. One finds that when we choose the infalling boundary condition for the scalar field near the outer horizon, the scalar field exhibits rapidly oscillations near the inner horizon under the synchronization condition \eqref{eq:syncon} resulting in a divergent energy-momentum tensor, which indicates that the inner horizon would be destroyed by the scalar field. Numerically, we indeed have not found any black hole with an inner horizon. Therefore it seems reasonable to expect that for $n\neq 0$ the hairy rotating black hole does not have any inner horizon.



\subsubsection{Collapse of the Einstein-Rosen bridge and oscillations}
\label{subs:er}
Similar to the previous case, inside the horizon there is an Einstein-Rosen bridge collapse close to the would-be-inner horizon. Different form the previous case, the scalar field oscillates after the collapse. This can be seen both analytically and numerically. 

When the system evolves inside the black hole while not far from the outer horizon, we can ignore the $(m^2+n^2z^2)\phi$ terms in the first and the last equations in \eqref{eq:eom3d2} and the right hand side of the second equation in \eqref{eq:eom3d2}\footnote{
Similar to the previous discussion in footnote \ref{ft:neglectscalar}, here we have also numerically checked that the $(m^2+n^2 z^2)\phi$ terms are approximately  on the order $10^{-3}$ in comparison to the other terms near the would-be inner horizon $z=z_*$.}. Then the equations of motion can be simplified as
\begin{align}\label{eq:eominner}
    \begin{split}
        N' &= E_0 z e^{-\chi/2}\,,\\
        z e^{\chi/2}\left(\frac{f e^{-\chi/2} \phi'}{z}\right)' &= -\phi\frac{e^{\chi}}{f} (\omega+ n N)^2\,,\\
        \frac{\chi'}{2z} &= \phi'^2 + \frac{e^{\chi}\phi^2}{f^2}(\omega+n N)^2\,,\\
        \left(\frac{e^{-\chi/2} f}{z^2}\right)' &= \frac{E_0^2}{2} z e^{-\chi/2} - 2\frac{e^{-\chi/2}}{z^3}\,.
    \end{split}
\end{align}

Similar to the $\omega=n=0$ case, near the would-be inner horizon $z=z_*$, we can set $z=z_*+\delta z$ and $f, \chi, N, \phi$ are functions of $\delta z$. The metric function $N$ is larger than its derivative, so we can set $N=N_0$ to be a constant in this regime. The equations of motion \eqref{eq:eominner} become
\begin{align}\label{eq:erc}
\begin{split}
    N &= N_0\,,\\
    e^{\chi/2}\left( f e^{-\chi/2} \phi' \right)' &= -\phi\frac{e^{\chi}}{f} (\omega+ n N_0)^2\,,\\
    \frac{\chi'}{2z_*} &= \phi'^2 + \frac{e^{\chi}\phi^2}{f^2}(\omega+n N_0)^2\,,\\
    \left(e^{-\chi/2} f\right)' &= \frac{E_0^2}{2} z_*^3 e^{-\chi/2} - 2\frac{e^{-\chi/2}}{z_*}\,.
    \end{split}
\end{align}
Solving the second equation above we obtain
\begin{align}\label{eq:solerphi}
    \phi= \varphi_0 \cos\left( (\omega+n N_0) \int_{z_*}^z  \frac{e^{\chi/2} }{f}\,d\tilde z\, + \varphi_1 \right)\,,
\end{align}
where $\varphi_0, \varphi_1$ are integration constants. 
Defining again $g_{tt0}=- f e^{-\chi}/z_*^2$, and taking the derivative of the fourth equation in  \eqref{eq:erc}, we obtain
\begin{align}
    \frac{g_{tt0}''}{g_{tt0}'}- \frac{c_1^2 g_{tt0}'}{g_{tt0}(c_1^2+g_{tt0})} =0, \quad \text{with} \quad c_1^2 = \frac{2\varphi_0^2 (\omega+nN_0)^2}{E_0^2 z_*^4-4}\,,
\end{align}
whose solution satisfies 
\begin{align}
\label{eq:solgtt}
    c_1^2 \log(g_{tt0}) + g_{tt0} = -c_2^2\,(\delta z + c_3)
\end{align}
where $c_2>0$ and $c_3$ are constants of integration. When $z<z_0$, $g_{tt0} \propto |\delta z|$ is linear vanishing, but when $z>z_0$, $g_{tt0} \propto e^{-(c_2/c_1)^2 \delta z}$ is exponentially decaying but always larger than $0$, indicating a collapse occurs 
over a coordinate range $\Delta z = (c_1/c_2)^2$. This behavior is quite similar to the previous case. 

\begin{figure}[h!]
\begin{center}
\includegraphics[width=0.46\textwidth]{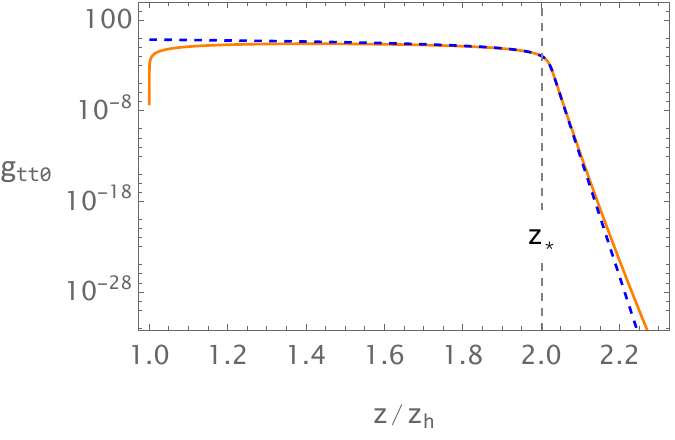}
~~~~
\includegraphics[width=0.45\textwidth]{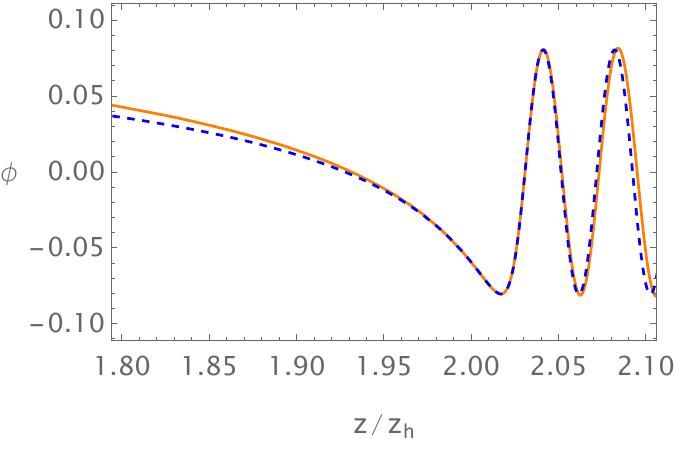}
\end{center}
\vspace{-0.3cm}
\caption{\small 
Plot of the Einstein-Rosen bridge collapse ({\em left}) and the scalar field $\phi$ near the would-be inner horizon ({\em right}). In both figures the solid orange line is our numerical solution of \eqref{eq:eombg}. The dashed blue line is the solution of \eqref{eq:solgtt} ({\em left}) and the numerical fitting of \eqref{eq:solerphi} ({\em right}) respectively. 
In this plot we have $\phi_0/\sqrt{J}=0.190\,, T/J=0.116\,,\omega/n=0.502$ and  $z_*=2.001\,,c_2/c_1=17.442$.}
\label{fig:ercomplex}
\end{figure}

An example of the collapse is shown in the left plot of Fig. \ref{fig:ercomplex}. 
This is similar to the case of $\omega=n=0$ shown in Fig. \ref{fig:ercneu}. The right plot in Fig. \ref{fig:ercomplex} shows that the scalar field  oscillates after the collapse 
and this is different from the case of $\omega=n=0$ where there is no oscillation for the scalar field $\phi$. 

From \eqref{eq:solgtt}, we have 
$
    g_{tt0}'=-\frac{c_2^2 g_{tt0}}{c_1^2+g_{tt0}}\,.
$
Then we can get the solution of $f,\chi$
\begin{align}
    f= -z_*^2 e^{\chi} g_{tt0}\,, \quad e^{-\chi} = \frac{c_2^2 g_{tt0}^2}{(c_1^2+g_{tt0})^2} \frac{2z_*^3}{E_0^2 z_*^4-4}\,.
\end{align}
Substituting them into \eqref{eq:solerphi}, we have 
\begin{align}
\label{eq:phiexpression}
    \phi=\varphi_0 \cos\left( \frac{c_1}{c_2\varphi_0} \frac{1}{\sqrt{z_*}} \log \frac{g_{tt0}(z)}{g_{tt0}(z_*)} + \varphi_1 \right)\,.
\end{align}
Note that the above equation only works close to $z_*$. The cosine behavior in \eqref{eq:phiexpression}  indicates the scalar field will start to oscillate after the collapse of Einstein-Rosen bridge, which is similar to holographic superconductor model \cite{Hartnoll:2020fhc}, while there is no cosine  behavior for the scalar field in the previous case. In the following we will show that the oscillation will continue for a longer regime. 

After the collapse of Einstein-Rosen bridge, the value of $e^{-\chi}$ becomes extremely small so that we can further simplify the first and last equations in \eqref{eq:eominner} as 
\begin{align}
\label{eq:eomphioss}
    \begin{split}
        N &= N_0\,,\\
        z e^{\chi/2}\left(\frac{f e^{-\chi/2} \phi'}{z}\right)' &= -\phi\frac{e^{\chi}}{f} (\omega+ n N_0)^2\,,\\
        \frac{\chi'}{2z} &= \phi'^2 + \frac{e^{\chi}\phi^2}{f^2}(\omega+n N_0)^2\,,\\
        \left(\frac{e^{-\chi/2} f}{z^2}\right)' &= 0\,.
    \end{split}
\end{align}
The last equation above implies 
\begin{align}\label{eq:fsoloss}
    \frac{e^{-\chi/2} f}{z^2}=-\frac{1}{c_4}
\end{align}
where $c_4$ is a constant of integration. Matching it with the $z>z_*$ solution of $g_{tt0}$ from \eqref{eq:solgtt} in the overlap region, we get
\begin{align}
    c_4=\frac{c_2}{c_1^2} \sqrt{\frac{2z_*^3}{E_0^2 z_*^4-4}}\,.
\end{align}

Plugging \eqref{eq:fsoloss} into the second equation of \eqref{eq:eomphioss}, we obtain the solution of scalar field
\begin{align}\label{eq:phiossi}
    \phi= c_5\, J_0 \left( \frac{|c_4(\omega+nN_0)|}{z} \right) + c_6\, Y_0 \left( \frac{|c_4(\omega+nN_0)|}{z} \right)
\end{align}
where $c_5, c_6$ are constants of integration. $J_0$ and $Y_0$ are Bessel functions which are oscillatory.

\begin{figure}[h!]
\begin{center}
\includegraphics[width=0.5\textwidth]{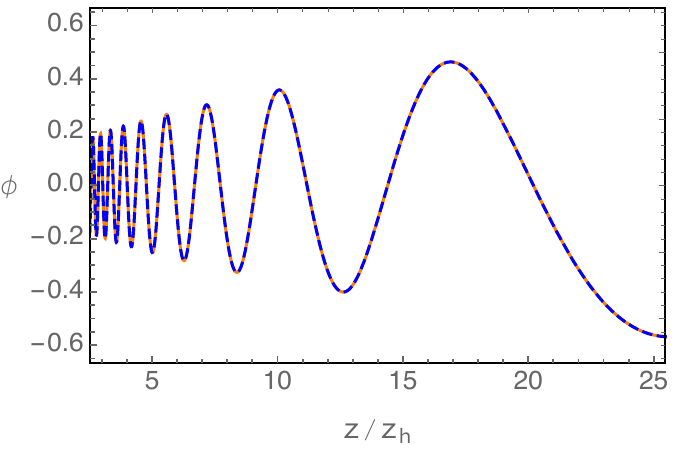}
\end{center}
\vspace{-0.3cm}
\caption{\small Plot of the scalar field $\phi$ for $\phi_0/\sqrt J=0.371\,, T/J=0.108\,,\omega/n=0.506$. The orange line is the numerical solution of \eqref{eq:eombg} and the dashed blue line is the fitting using \eqref{eq:phiossi}.}
\label{fig:phiossfit}
\end{figure}

Fig. \ref{fig:phiossfit} shows a typical example of the oscillation of the scalar field $\phi$. From the figure, we can see that the full numerical solution of \eqref{eq:eombg} and the approximate analytical solution \eqref{eq:phiossi} match very well. We can also see that the frequency of oscillation decreases and the amplitude increases as $z/z_h$ increases.  

\begin{figure}[h!]
\begin{center}
\includegraphics[width=0.31\textwidth]{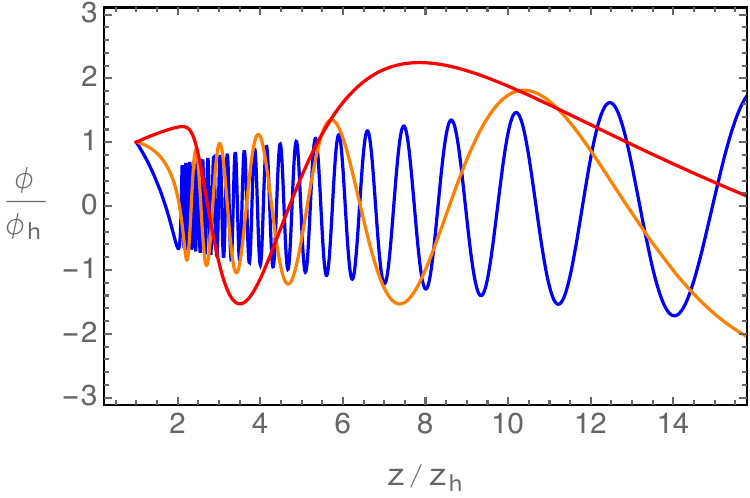}
~
\includegraphics[width=0.31\textwidth]{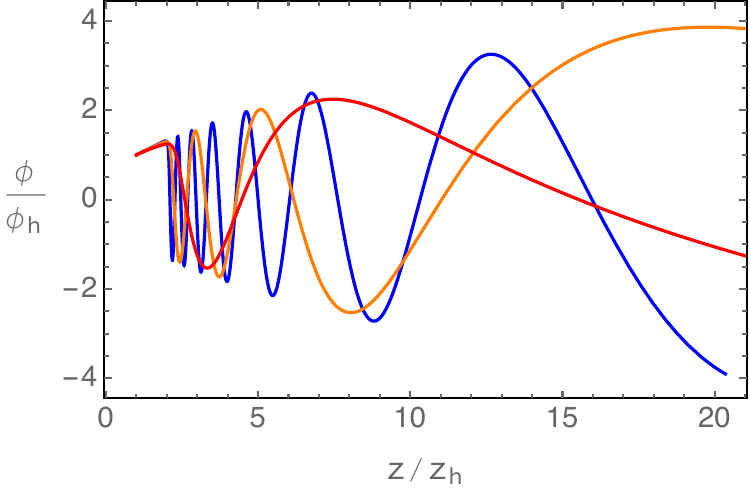}
~
\includegraphics[width=0.32\textwidth]{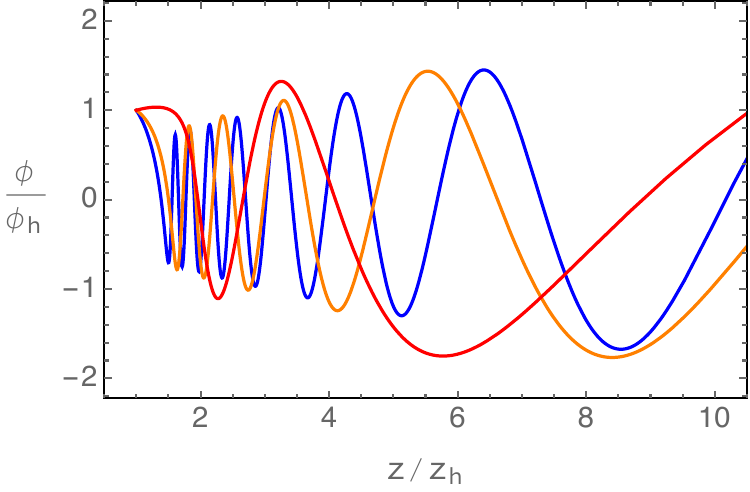}
\end{center}
\vspace{-0.3cm}
\caption{\small {\em Left:} Plots of the scalar field for $\phi_0/\sqrt{J}=0.5\,, \omega/n=0.5$ and $T/J=0.15$ (blue), $0.1$ (orange) and $0.05$ (red). {\em Middle:} Plots of the scalar field for $T/J=0.05\,, \omega/n=0.5$ and $\phi_0/\sqrt{J}=0.2$ (blue), $0.3$ (orange) and $0.5$ (red). {\em Right:} Plots of the scalar field for $T/J=0.05\,, \phi_0/\sqrt{J}=0.5$ and $\omega/n=0.73$  (blue), $0.7$ (orange) and $0.6$ (red). }
\label{fig:ossvar}
\end{figure}

In Fig. \ref{fig:ossvar}, we show the scalar field $\phi$ as a function of $z/z_h$ for fixed $\phi_0/\sqrt{J}, \omega/n$ (left) or $T/J, \omega/n$ (middle) or $T/J, \phi_0/\sqrt{J}$ (right) respectively in the oscillation regime. The left plot in Fig. \ref{fig:ossvar} shows that for fixed $\phi_0/\sqrt{J}$ and $\omega/n$, the field $\phi$ oscillates less dramatic at lower temperature which is similar to the holographic superconductor model \cite{Hartnoll:2020fhc}. This is also consistent with 
the discussion in \cite{Dias:2019ery} that in the probe limit the scalar field is more smooth at higher temperature. The middle plot shows that for fixing temperature $T/J$ and $\omega/n$, the smaller $\phi_0/\sqrt{J}$ is, the more times $\phi$ oscillates, and the right plot shows that for fixing temperature $T/J$ and $\phi_0/\sqrt{J}$ the frequency of the oscillation is larger for larger $\omega/n$.  

The oscillation of scalar field also backreacts the metric field $f$ and $\chi$. Inserting \eqref{eq:phiossi} into the third equation of \eqref{eq:eomphioss}, we get the solution 
\begin{align}
    \chi &= 2\log(f_0 c_4) + \int_{z_*}^z \left[ \frac{2c_4^2(\omega + nN_0)^2\phi^2}{\tilde z^3} + 2\tilde z \phi'^2 \right] d\tilde z\,,\\
    f &= -f_0 z^2 \exp \left\{\int_{z_*}^z \left[ \frac{2c_4^2(\omega + nN_0)^2\phi^2}{\tilde z^3} + 2\tilde z \phi'^2 \right] d\tilde z \right\} \,,
\end{align}
where $f_0$ is an integration constant. 

At large $z$, by expanding \eqref{eq:phiossi} at $z\to\infty$, the behavior of scalar field  becomes
\begin{align}
\label{eq:philog}
    \phi \sim \frac{2c_6}{\pi} \log\left( \frac{c_4(\omega+nN_0)}{2z} \right)+ c_5 + \frac{2c_6 \gamma_E}{\pi} + \cdots
\end{align}
where $\gamma_E$ is the Euler constant. The logarithmic behavior is a typical behavior of the scalar field near the Kasner singularity.

\subsubsection{Kasner inversion and Kasner singularity}
\label{sec3}

After the oscillation regime of the scalar field (if it exists), the system evolves into the Kasner epoch. Different from the previous case, the Kasner inversion might exist. In the following we first analytically describe the Kasner epoch and  the Kasner inversion and then provide numerical analysis. 

As we have already shown in \eqref{eq:philog}, the system will evolve into the Kasner regime in the deep interior of the black hole. The simplified equations near the singularity $z\rightarrow \infty$ are nearly the same with $\omega=n=0$ case \eqref{eq:simplified eqs} except that the term proportional to $N'^2$ can not be ignored in the last equation which triggers the Kasner inversion 
\begin{align}\label{eq:eomsincom}
    \begin{split}
         z e^{\chi/2}\left(\frac{f e^{-\chi/2} \phi'}{z}\right)' &= 0 \,,\\
       z e^{-\chi/2} \left(\frac{e^{\chi/2} N'}{z}\right)' &= 0\,,\\
       \frac{\chi'}{2z} &= \phi'^2  \,,\\
       2e^{\chi/2}z^3 \left(\frac{e^{-\chi/2} f}{z^2}\right)' &= z^2 e^{\chi} N'^2\,.
    \end{split}
\end{align}

Taking the derivative of the first equation and eliminating $f'', \chi', f'$ and $f$ from the other three equations, we obtain a third order ODE of the scalar field $\phi$
\begin{align}\label{eq:3rdphi}
    \phi'''-\frac{2\phi''^2}{\phi'} + \left( z\phi'^2-\frac{3}{z} \right) \phi'' +\phi'^3 - \frac{3\phi'}{z^2} = 0\,.
\end{align}
For analyzing the behavior of $\phi$, we set
\begin{align}\label{eq:phitos}
    \phi=\int \frac{\tilde{\alpha}(\phi)}{z} dz
\end{align}
then \eqref{eq:3rdphi} becomes a second order ODE of $\tilde \alpha(\phi)$
\begin{align}
\label{eq:tildealpha}
   \ddot{\tilde{\alpha}} - \frac{{\dot{\tilde{\alpha}}}^2}{\tilde{\alpha}} + \tilde{\alpha}\dot{\tilde{\alpha}} - \frac{2\dot{\tilde{\alpha}}}{\tilde{\alpha}}  = 0\,,
\end{align}
where the dot is the derivative with respect to $\phi$. 
The above equation \eqref{eq:tildealpha} has an analytical solution 
\begin{align}
\label{eq:tildea}
    \tilde{\alpha} = \frac{1}{2} \left( \beta_i + \sqrt{\beta_i^2-8}\, \tanh\left[ \frac{\sqrt{\beta_i^2-8}\, (\phi-\phi_i)}{2} \right]  \right)
\end{align}
where $\beta_i, \phi_i$ are two constants of integration and we have assumed $|\beta_i|> 2 \sqrt 2$. The solution \eqref{eq:tildea} has the limit behavior
\begin{align}
\label{eq:inversion1}
    \tilde{\alpha}\rightarrow \frac{\beta_i+\sqrt{\beta_i^2-8}}{2} \qquad &\text{as} \qquad  \phi \gg \phi_i\,,\\
    \label{eq:inversion2}
    \tilde{\alpha}\rightarrow \frac{\beta_i-\sqrt{\beta_i^2-8}}{2} \qquad &\text{as} \qquad  \phi \ll \phi_i\,.
\end{align}
Obviously $\phi_i$ is the location where $\tilde\alpha$ jumps. As can be seen from \eqref{eq:phitos}, different limits \eqref{eq:inversion1} and \eqref{eq:inversion2} correspond to different Kasner epoches. Therefore when  $\phi$ is close to $\phi_i$, the Kanser inversion  happens. From \eqref{eq:inversion1} and \eqref{eq:inversion2}, we know that $\tilde\alpha$ changes before and after the Kasner inversion in the form 
\begin{align}\label{eq:alphachange}
    \tilde\alpha \to \frac{2}{\tilde\alpha}\,.
\end{align}
We will call the Kasner epoch before inversion intermediate Kasner epoch and the one after inversion final Kasner epoch.

\begin{figure}[h!]
\begin{center}
\includegraphics[width=0.5\textwidth]{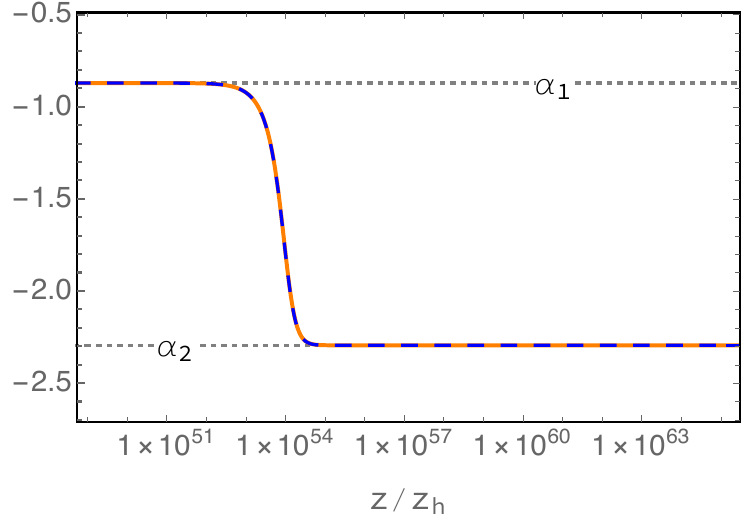}
\end{center}
\vspace{-0.3cm}
\caption{\small Plot of $\tilde{\alpha}=z \phi'$ as a function of $z/z_h$ when $\phi_0/\sqrt J=0.371\,, T/J=0.108\,,\omega/n=0.506$. The orange line is the numerical solution of \eqref{eq:eom3d2}, while the dashed blue line is the solution of \eqref{eq:3rdphi}. Note that $\alpha_1=-0.872$ in the intermediate Kasner epoch and $\alpha_2=-2.295$ in the final Kasner epoch satisfying $\alpha_1 \alpha_2=2$. 
}
\label{fig:alphafit}
\end{figure}

A typical example of Kasner inversion is shown in Fig. \ref{fig:alphafit}. In the intermediate and final Kasner epoch, $\tilde{\alpha}$ are constants of $\alpha_1$ and $\alpha_2$ respectively. The orange line is from numerics of the solution while the dashed blue line is the solution of \eqref{eq:3rdphi}. 
The perfectly matching of these two lines indicates that  \eqref{eq:3rdphi} is a good approximation of the entire system near singularity. The time domain for inversion is extremely short comparing to the Kasner epoches. 

At each Kasner epoch, we set $\tilde\alpha(\phi)=\alpha$ to be a constant and the solution of \eqref{eq:eomsincom} is the same as  $\omega=n=0$ case with the following form 
\begin{align}
\label{eq:NN}
    f \sim -f_K z^{2+\alpha^2}\,, ~~~~\chi \sim 2\alpha^2 \log z + \chi_1 \,, ~~~~\phi \sim \alpha \log z\,,~~~~
    N\sim N_K+ \frac{E_K z^{2-\alpha^2}}{2-\alpha^2}\,.
\end{align}
If $\alpha^2<2$ in the intermediate Kasner epoch, then  the right hand side in the last equation of \eqref{eq:eomsincom} becomes more and more important when we increase $z$, resulting in a possible  instability for  the above Kasner form. In this case Kasner inversion occurs and the geometry further flows into a new stable Kasner epoch and we will have $\alpha^2>2$ in the end. Note that the Kasner inversion may not exist when it evolves directly into a Kasner singularity with $\alpha^2>2$ or it takes extremely short time to evolve into a Kasner singularity with $\alpha^2>2$ so that the intermediate Kasner epoch can not exist. Also note that there is only one inversion at most because $(2/\alpha)^2>2$ for any $\alpha^2<2$.   

\begin{figure}[h!]
\begin{center}
\includegraphics[width=0.31\textwidth]{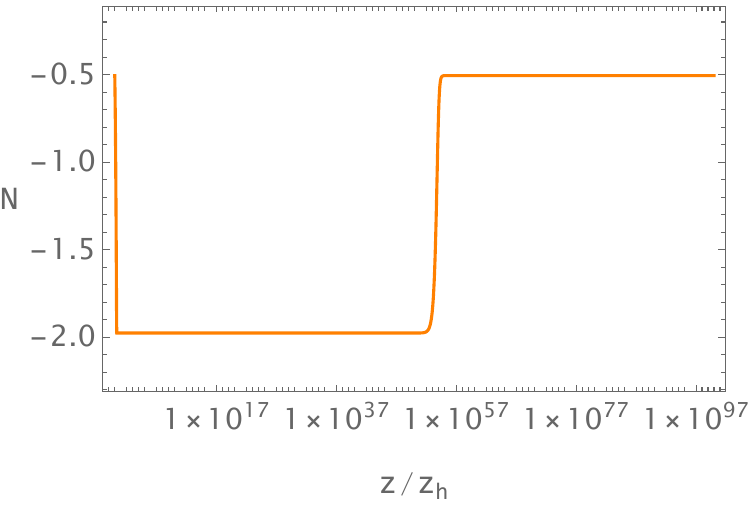}
~~
\includegraphics[width=0.318\textwidth]{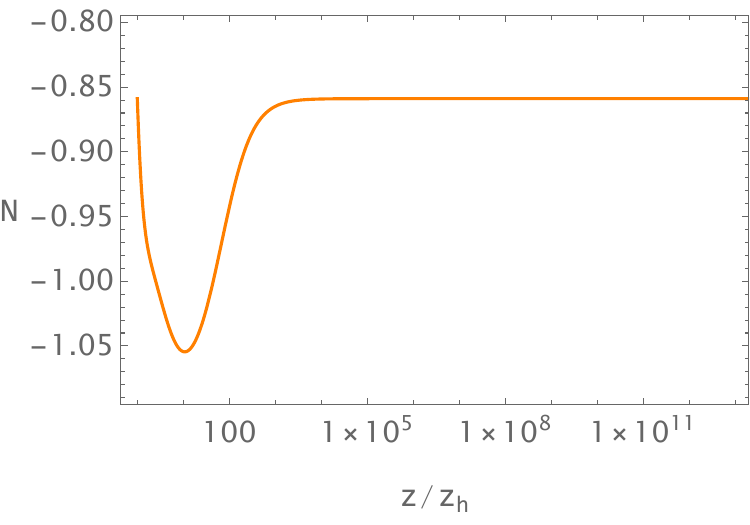}
~~
\includegraphics[width=0.31\textwidth]{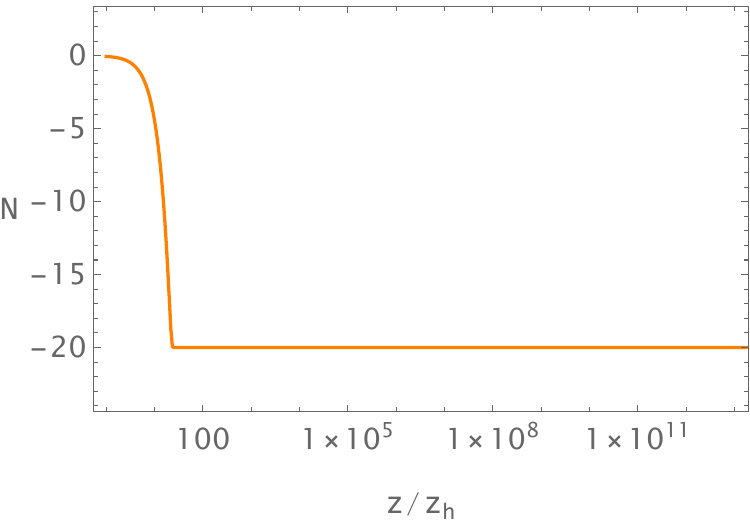}
\end{center}
\vspace{-0.3cm}
\caption{\small  Plot of $N$ as a function of $z/z_h$ for $\phi_0/\sqrt J=0.371\,, T/J=0.108\,,\omega/n=0.506$ ({\em left}), $\phi_0/\sqrt J=1\,, T/J=0.005\,,\omega/n=0.859$ ({\em middle}) and $\phi_0/\sqrt J=0.5\,, T/J=0.05\,,\omega/n=0.05$ ({\em right}).
}
\label{fig:inveN}
\end{figure}

Three typical interior configurations of the metric field $N$ are shown in Fig. \ref{fig:inveN}. 
We find that $N$ is a monotonic function of $z$ before the Kasner inversion.  
All these three plots show that $N$ can be treated as a constant in both Kasner epochs. The left plot also indicates the value of $N$ at the horizon $N_h$ is the same as that at the final singularity $N_K^{(f)}$ when the Kasner inversion exists, i.e. $N_K^{(f)}=N_h$. 
In the intermediate Kasner regime, we have $N_K^{(i)}=1/N_h$.\footnote{It would be interesting to obtain them analytically and to  check if these relations still hold  for higher dimensional hairy rotating black holes.} 
We have verified that the above two relations are always  true whenever there is a Kasner inversion.  Moreover, numerically we find that when there is no Kasner inversion, we have $N_K=1/N_h$, which is the same as the previous case of real scalar field. Note that the phase velocity of the scalar field is related to the horizon value of $N$ \eqref{eq:syncon}. In this way we see that the source on the boundary is connected to the singularity.

From the fact that $N$ is almost a constant during each Kasner epoch as shown in Fig. \ref{fig:inveN}, we can perform a coordinate transformation
\begin{align}
    \tau = \frac{2}{\sqrt{f_K}(\alpha^2+2)}\, z^{-(\alpha^2+2)/2}
\end{align}
and obtain the Kasner form for the fields
\begin{align}
\label{eq:kasnerform}
\begin{split}
    ds^2 &= -d\tau^2 + c_t \tau^{2p_t} dt^2 + c_x \tau^{2p_x} (Ndt+dx)^2\,,\\
    \phi &=p_{\phi}\log\tau + c_{\phi}\,,
\end{split}
\end{align}
where
\begin{align}\label{eq:kasnerexp}
    p_t = \frac{\alpha^2}{\alpha^2+2}\,,~~~ p_x= \frac{2}{\alpha^2+2}\,,~~~~ p_{\phi} = -\frac{2\alpha}{\alpha^2+2}\,.
\end{align}
Obviously, the Kasner relations $p_t+p_x=p_t^2+p_x^2+p_\phi^2=1$ are satisfied.

\begin{figure}[h!]
\begin{center}
\includegraphics[width=0.5\textwidth]{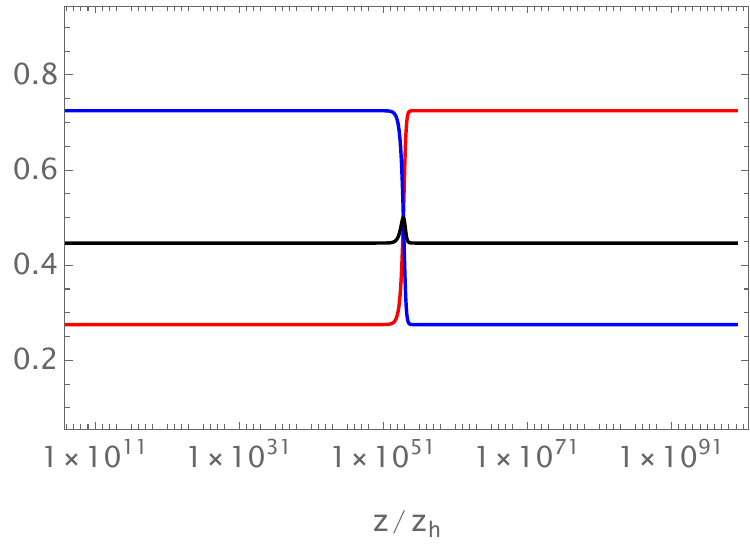}
\end{center}
\vspace{-0.3cm}
\caption{\small  Plots of Kasner exponents 
$p_t$ (red)\,,~~$p_x$ (blue)\,and $p_{\phi}$ (black) 
as a function of $z/z_h$ when $\phi_0/\sqrt J=0.371\,, T/J=0.108\,,\omega/n=0.506$. The Kasner exponents before inversion are $p_t=0.275, p_x=0.725, p_{\phi}=0.447$ and after inversion are $p_t=0.725, p_x=0.275, p_{\phi}=0.447$. 
}
\label{fig:inver}
\end{figure}

When the Kasner inversion ocurrs, from  \eqref{eq:alphachange} and \eqref{eq:kasnerexp}, we have the following transformations of the Kasner exponents before and after the Kasner inversion:   
\begin{align}
\label{eq:kasnerinv}
    p_t \rightarrow p_x\,,~~~ p_x \rightarrow p_t\,,~~~ p_{\phi} \rightarrow p_{\phi}\,.
\end{align}
A typical example for the Kanser exponents during the inversion is shown in Fig. \ref{fig:inver}.

\begin{figure}[h!]
\begin{center}
\includegraphics[width=0.313\textwidth]{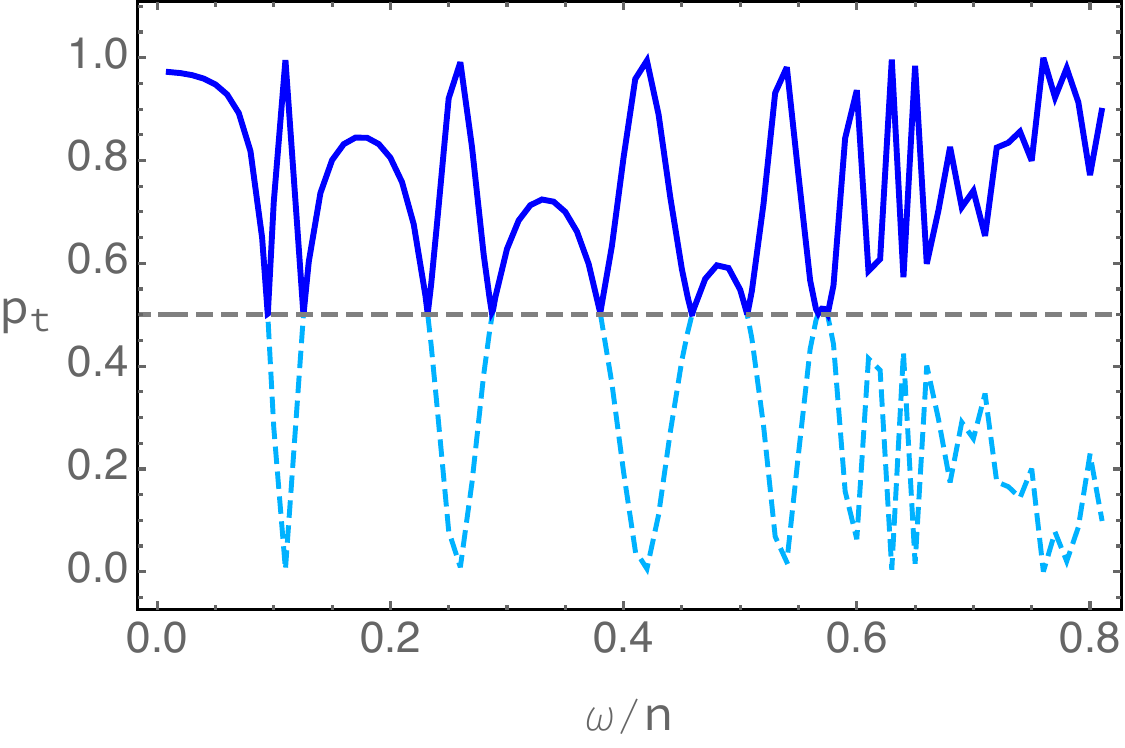}
~
\includegraphics[width=0.313\textwidth]{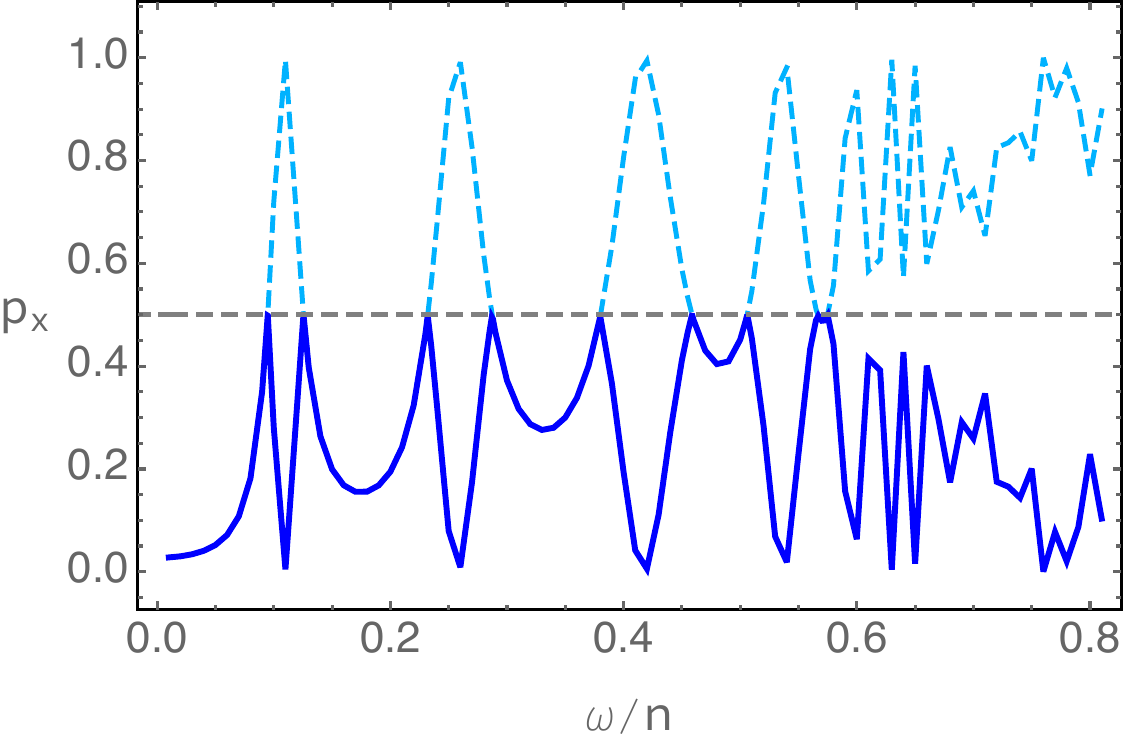}
~
\includegraphics[width=0.317\textwidth]{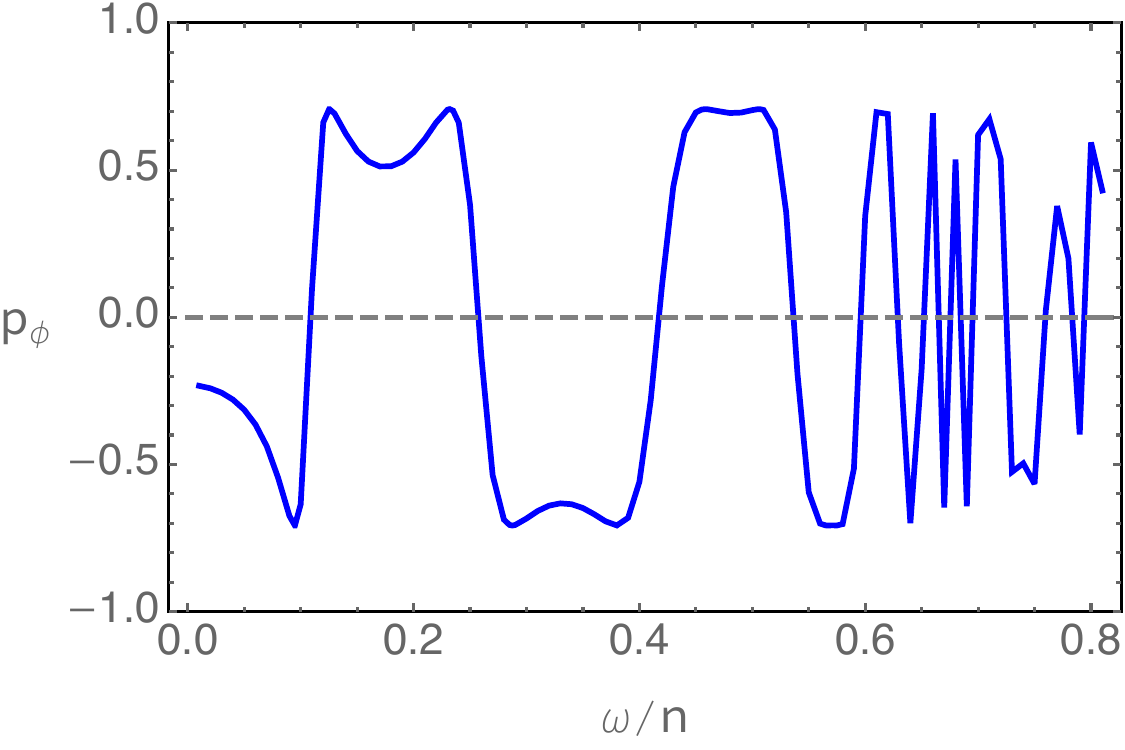}
\end{center}
\vspace{-0.3cm}
\caption{\small The Kasner exponents as functions of $\omega/n$ for fixed $\phi_0/\sqrt J=0.5, T/J=0.05$. The dashed light blue lines are the intermediate Kasner exponents, while the solid blue lines correspond to final Kasner exponents. 
}
\label{fig:ptcha}
\end{figure}

Finally, we show the final Kasner exponents as functions of $\omega/n$ for fixing $\phi_0/\sqrt{J}$ and $T/J$ in Fig. \ref{fig:ptcha}. The solid blue lines are for the final Kasner exponents while the dashed light blue lines are for the intermediate Kanser exponents. For $p_\phi$ the two lines coincide. We have verified that the Kasner inversion relation \eqref{eq:kasnerinv} is always satisfied. Near the regime $\omega/n\to 1$, the Kanser inversion always occurs and the Kanser exponents are sensitive to the parameter. When $\omega/n$ is small there are regimes that we do not have Kasner inversion. Different from the black holes in case I where typically the Kasner exponents are smooth functions of the external parameters, here these functions are not smooth at certain locations of $p_t=p_x=0.5, p_\phi=\pm\sqrt{2}/2$. In most parts of the parameter regimes, the Kasner inversion occurs and the behaviors of Kasner exponents become irregular, which is different from that in \cite{Hartnoll:2020rwq}. 

\section{Conclusions and discussions}
\label{sec:cd}

In this work we have studied two kinds of hairy rotating black holes in three dimensional Einstein gravity coupled to a complex scalar field. In the first case,  the scalar field is real and the system has two Killing vectors. We find that there is no inner horizon in this case. The system evolves smoothly into a stable Kasner regime. In the second case, the rotating black hole with a complex scalar field has only one Killing vector. We find that inside the black hole, after a collapse of the Einstein-Rosen bridge, the scalar field can oscillate dramatically, then the system evolves into Kasner epoch. It turns out that at lower temperatures the scalar field oscillates fewer times. This behavior shares similarity with the properties of the scalar field close to the Cauchy horizon of the rotating BTZ in the probe limit discussed in \cite{Dias:2019ery}. Moreover, there could be Kasner inversion for some parameter regimes. Different from the interior of holographic superconductors  \cite{Hartnoll:2020fhc}, in most of the parameter regimes, the Kasner inversion occurs. In both these two cases, we find an interesting simple relation between the metric field $N$ at the horizon and the value of $N$ (to the leading order) in the intermediate or final Kanser epoch.  Our study shows that the internal structure of certain stationary black holes and time-dependent black holes behaves similarly to the static black holes, indicating a possible universality of the internal structure of black holes. 

For nonzero $\omega$, the time shift symmetry is explicitly broken into $t\to t+\frac{2\pi}{\omega}$ for the hairy black hole, which leads to an energy dissipation in the dual system. Further exploration of the dual field is necessary. For example, it should be interesting to connect it to the time crystal. It is also  interesting to study the quantum chaos for the dual field theory from the method of OTOC and pole skipping to see if they give the same result \cite{Grozdanov:2017ajz, Blake:2018leo}. In this way one could understand better the relationship between hydrodynamics and quantum chaos. 

It would be interesting to study if the hairy black hole solution studied in this work could serve as a final state during the gravitational collapse in three dimensional Einstein-scalar theory \cite{Pretorius:2000yu, Pandya:2020ejc}. In order to explore this question, it is clear that one needs to consider a fully dynamical process under a generic initial condition to study the final state. 

We have only studied the three dimensional hairy rotating black holes. A qualitatively similar internal structure has been seen in four dimensional asymptotic flat hairy rotating black hole \cite{Brihaye:2016vkv}.  It is interesting to consider other higher dimensional hairy rotating black holes to study their interior geometries \cite{Dias:2010ma, Kim:2023sig}. We expect that such exploration will help us further understanding the universal physics of the internal structure of black holes. 

\subsection*{Acknowledgments}
We thank C. Herdeiro for drawing our attention to \cite{Brihaye:2016vkv}. 
This work is supported by the National Natural Science Foundation of China grant No. 11875083,
12375041.

\appendix

\vspace{.5cm}
\section{The low temperature solution}
\label{sec:lt}

In this appendix, we show the low temperature solution of the two cases discussed in Sec. \ref{subsec:cI} and Sec. \ref{subsec:cII}. At low temperatures, the solutions behave quite interestingly outside the horizon. The profile of the metric field $f$ is no longer monotonic. On the other hand, we have not any special behavior inside the horizon for black holes at low temperatures. 

\begin{figure}[htbp]
\begin{center}
\includegraphics[width=0.31\textwidth]{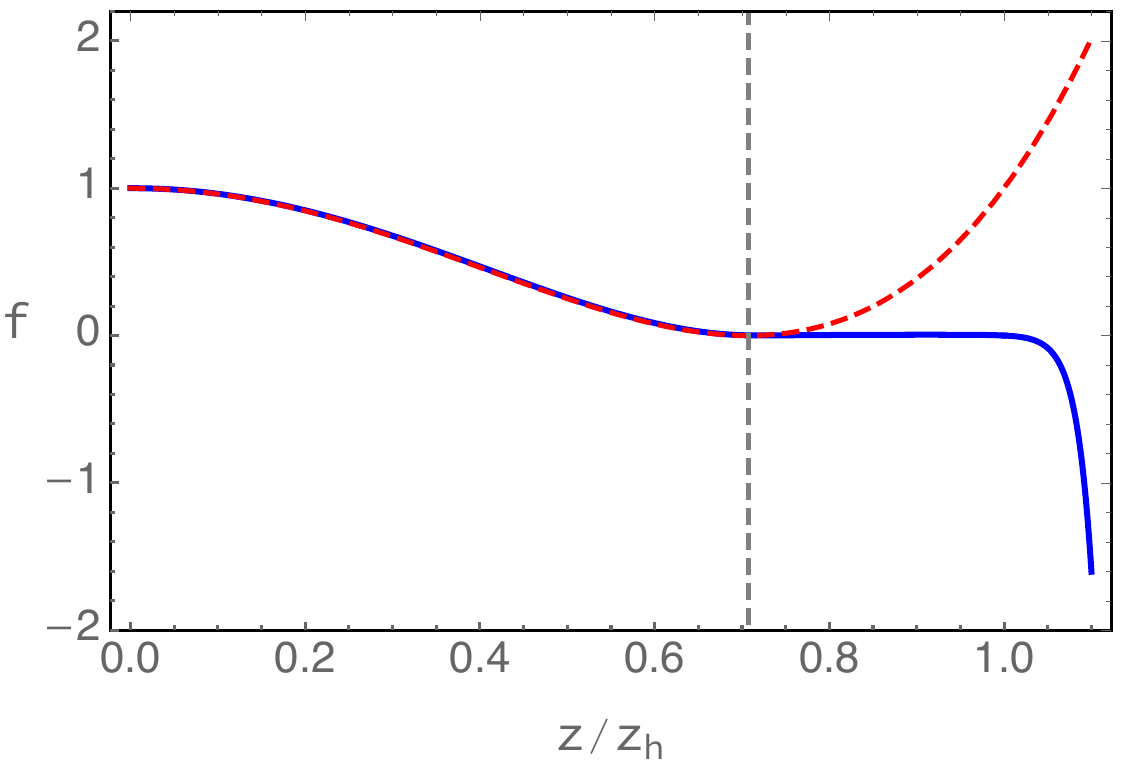}~~~
\includegraphics[width=0.32\textwidth]{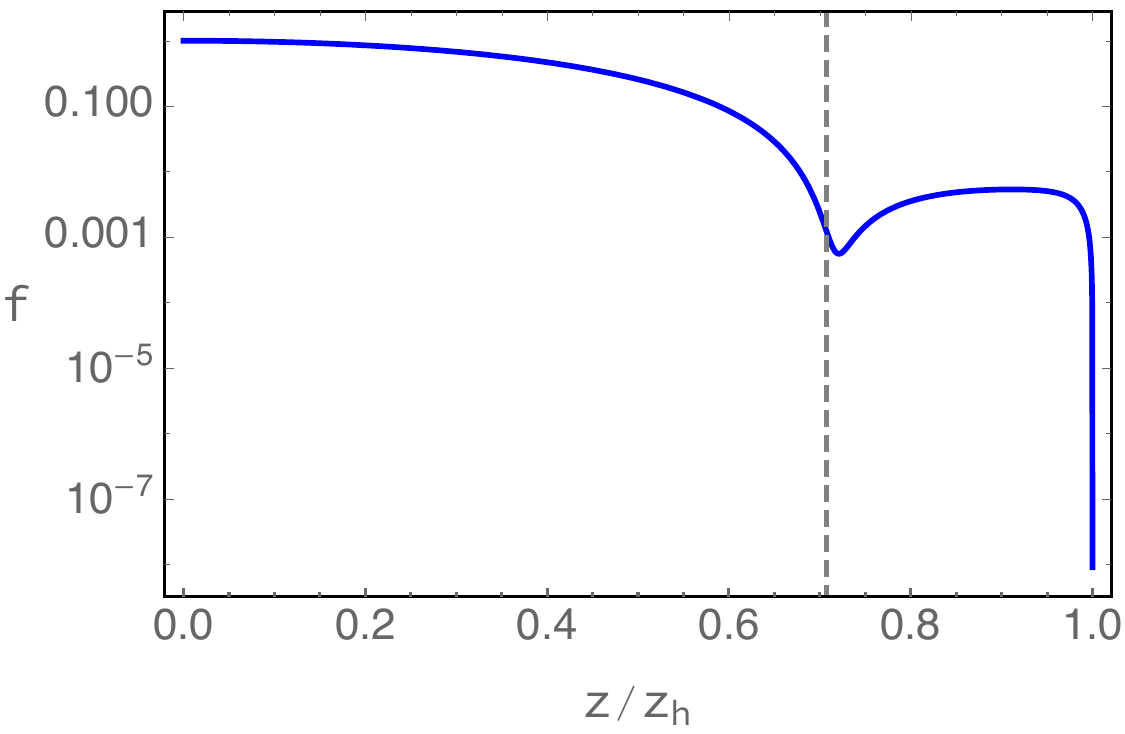}~~~
\includegraphics[width=0.34\textwidth]{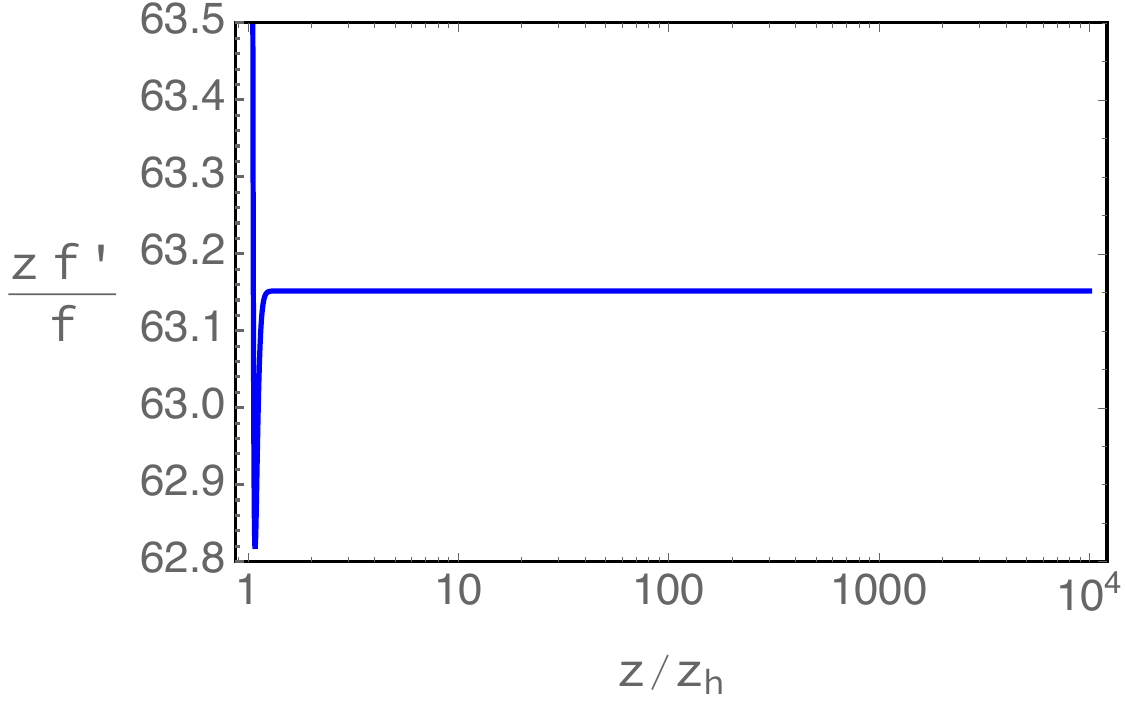}

\end{center}
\vspace{-0.3cm}
\caption{\small  Plots of the metric field $f$ for $T/J=4\times 10^{-11}$ and $ \phi_0/\sqrt{J}=0.103$. The blue lines are the numerical solutions of hairy rotating black holes, red dashed lines are solutions of the extremal BTZ black hole with the ``horizon'' $z_0$, and gray dashed line is the location of the ``horizon''. The metric field $f$ has a minimum value near $z=z_0$ ($z_0<z_h$).
}
\label{fig:real-extreme}
\end{figure}

More precisely, for the case I of $\omega=n=0$, i.e. the real scalar field, the function $f$ has a minimum value in the vicinity of $z_0$ which is located between the horizon and boundary, see Fig. \ref{fig:real-extreme}. This happens when the temperature is quite low at any value of  $\phi_0/\sqrt{J}$, i.e. $T/J\to 0$. In this case we find that the angular velocity of the black hole approaches the speed of light.\footnote{Our numerics works at finite temperature. At zero temperature, we should use another near horizon expansion to get the full solution.} Even though we have chosen the event horizon to be $z_h$, $z_0$ can be regarded as the ``horizon" of an extremal BTZ black hole whose spacetime outside the ``horizon" match very well with this extremal hairy black hole. Moreover, we have $J/M \rightarrow 1$ which implies that we reach the upper limit of the angular momentum for hairy rotating black holes. 

\begin{figure}[h]
\begin{center}
\includegraphics[width=0.31\textwidth]{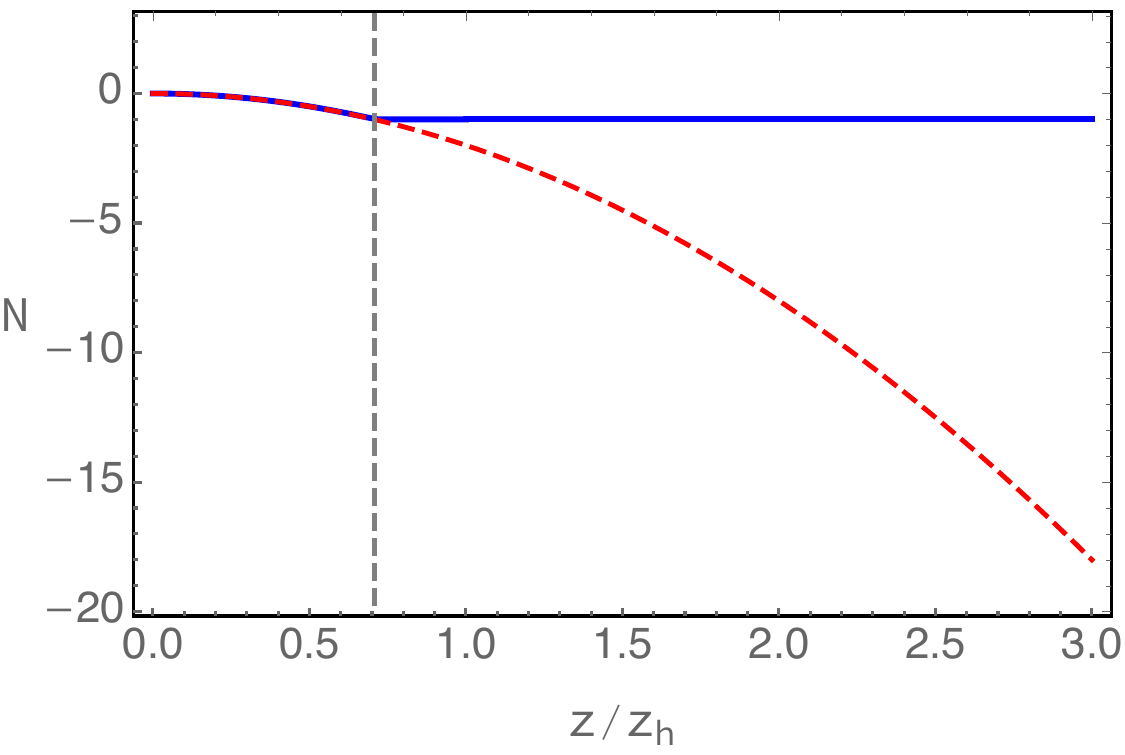}~
\includegraphics[width=0.315\textwidth]{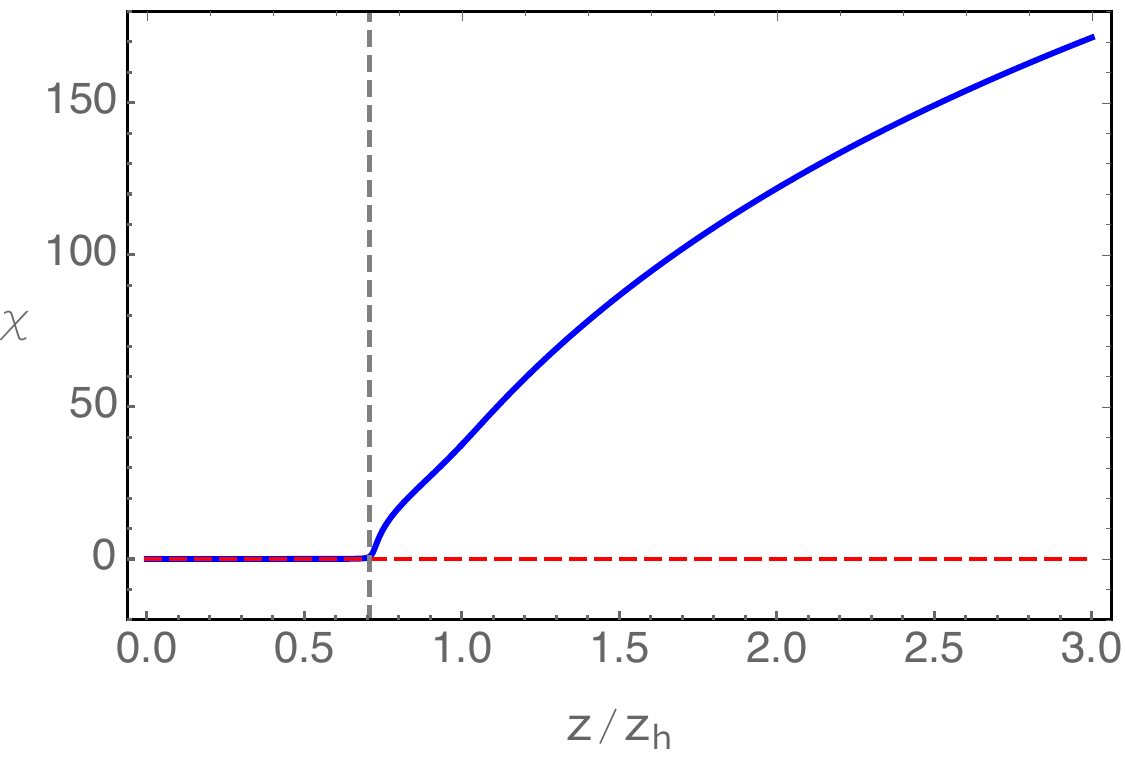}~~
\includegraphics[width=0.31\textwidth]{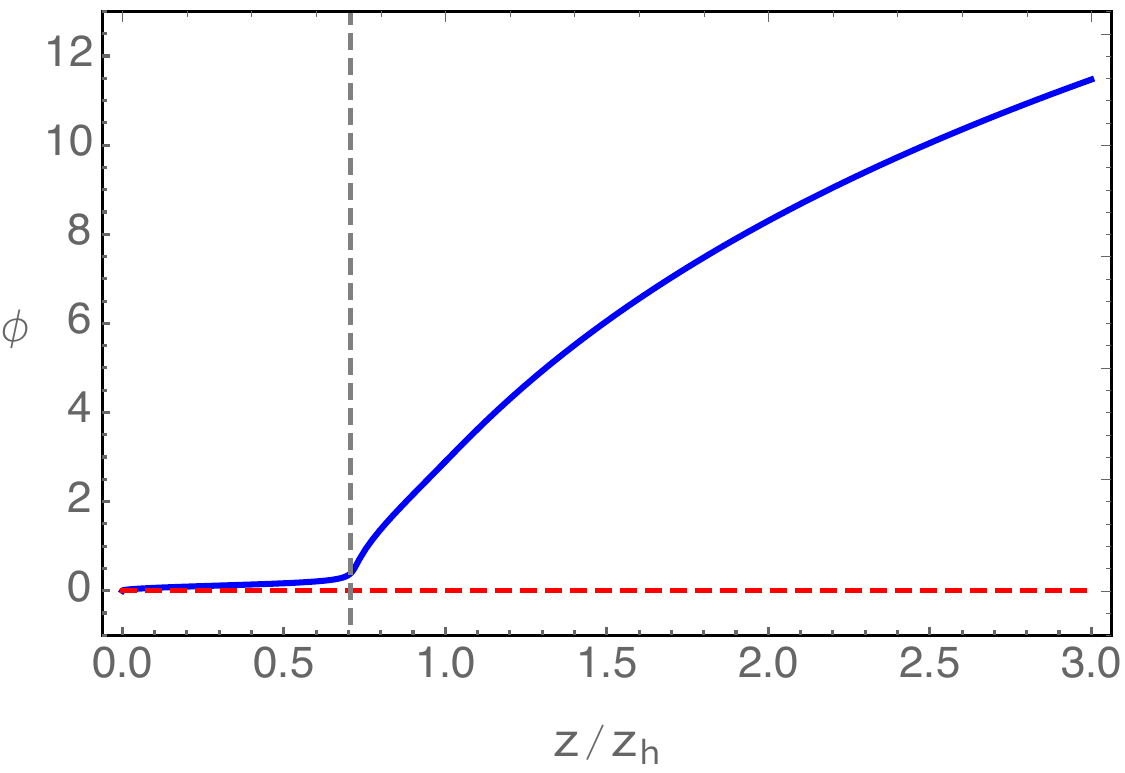}~~~
\end{center}
\vspace{-0.3cm}
\caption{\small Plots of the fields $N$, $\phi$ and $\chi$ for $T/J=4\times 10^{-11}$ and $ \phi_0/\sqrt{J}=0.103$. The blue lines, red dashed lines and gray dashed lines are the numerical solutions, solutions of the extremal BTZ and the location of the ``horizon'' respectively. }
\label{fig:real-extreme2}
\end{figure}

Inside the hairy black hole at low temperatures, it behaves differently compared with the extremal BTZ black hole, as shown in Fig. \ref{fig:real-extreme2}. Now there is a curvature singularity of Kasner form inside the hairy black hole, taking the Kasner form $f\sim-0.004 \,z^{63.15},\,\chi\sim 122.3 \log z,\,\phi\sim 7.82 \log z,\,N\sim-1 $, while the singularity of extremal BTZ is a conical one, i.e. $f\sim z^4/z_0^4,\,N\sim -z^2/z_0^2,\,\chi=\phi=0$. 
Moreover, the scalar field outside the horizon mainly stays in between $z_0$ and $z_h$. 

For the case of $n\neq 0$, i.e. the complex scalar field, we have three parameters for the hairy black holes. It turns out that the above behavior happens only at low temperature $T/J\to 0$ when the parameter $\omega/n \rightarrow 1$. In this case the angular velocity of the black hole is close to the speed of light. 
Inside the hairy black hole, we find that there is no oscillation of the scalar field and no Kasner inversion. So the black hole interiors are also similar to the case of a real scalar field.    

It is natural to suspect that the low temperature solutions are connected to the branch of the boson stars at zero temperature. In other words, when the mass $M/J$ of the boson star at zero temperature approaches a critical value, the extremal BTZ black hole appears and the system evolves to the hairy solution discussed above. Similar behavior has been discussed in  \cite{Stotyn:2013spa}.

\section{The non-existence of the inner horizon in case II}
\label{app:nonih}

In this appendix we first show that the non-existence of the  inner horizon can not be proved analytically using the methods developed in recent literature, see e.g. \cite{Hartnoll:2020rwq, Hartnoll:2020fhc, Cai:2020wrp, An:2022lvo, Liu:2022rsy, Gao:2023zbd, Cai:2021obq, Sword:2022oyg, Mansoori:2021wxf, Mirjalali:2022wrg, Grandi:2021ajl, Devecioglu:2023hmn, Dias:2021afz}. Then we show that in the probe limit the backreaction of the scalar field is expected to destroy the inner horizon of the BTZ black hole, which is consistent with our numerical results.

We start from the method widely used in literature for proving the no inner horizon and show that they fail. The idea is to find an inconsistent relation with the assumption of the existence of more than one event horizon, e.g. two event horizons $z_h$ and $z_i$. 

The first method is to obtain relations using the equations of motion \cite{Hartnoll:2020rwq}.  
From the first equation of \eqref{eq:eom3d2} we have
\begin{align}
    0= \int_{z_h}^{z_i} \left( \frac{fe^{-\chi/2}\phi\phi'}{z} \right)' dz = \int_{z_h}^{z_i} \frac{e^{-\chi/2}}{z^3} \left[ z^2 f \phi'^2+ \phi^2 \left( m^2+n^2z^2 - \frac{e^{\chi}}{f}z^2(\omega+nN)^2 \right) \right] dz\,.
\end{align}
The first equality is because of  $f(z_h) = f(z_i) = 0 $. For the right hand side, although the first term and second term in the integrand are negative because we have $f(z) < 0$
between the two horizons and $m^2 < 0$ but the last two terms are positive so that there is no conflict. 
Another equality we find from the equations of motion is 
\bea
\int dz\, \big(z^2 f\phi'^2+2\big) \frac{e^{-\chi/2}}{z^3}=0\,
\eea
from which the nonexistence of the inner horizon can not be verified. 

The second method is to use conserved quantity \cite{Hartnoll:2020fhc, Cai:2020wrp}.  
We find there is a radial conserved charge in our system 
\be
Q(z)=\frac{1}{z}e^{\chi/2}(f e^{-\chi})'-\frac{e^{\chi/2}}{nz}(\omega +n N) N'-\frac{2f}{z^2}e^{-\chi/2}-\int_{z_h}^z dx \frac{e^{-\chi/2}}{x^3}(4-2m^2\phi^2)
\ee
and it can be checked $Q'=0$ by using equations of motion \eqref{eq:eombg}. Assuming there are two horizons $z_h$ and $z_i$ and evaluating $Q$ at the two horizons we have 
\be\label{eq:ts}
T_hS_h+T_iS_i=2 \pi\int_{z_h}^{z_i}{\mathrm{d}z}\, \frac{e^{-\chi/2}}{z^3}(4-2 m^2\phi^2)
\ee
where $T_{h,i}$ and $S_{h,i}$ are the temperature and thermal entropy at the outer and inner horizon. Note that \eqref{eq:ts} are positive on both side so it can not be used to eliminate the inner horizon. 

The third method is essentially the same as the second one, i.e. using the Killing vector \cite{{Dias:2021afz}}. Assume that there are two horizons, which are both expected to be Killing horizons with generators $\xi^{h,i}=\partial_t+\Omega_{h,i}\, \partial_x$. In order to guarantee the scalar $\phi$ to be invariant at the two horizons, we must take $\Omega_{h}=\Omega_{i}=\Omega=\frac{\omega}{n}$ and thus $\xi^h = \xi^i = \xi$. Then, we can integrate some quantity associated with the only Killing vector $\xi$ to get a contradiction.

More precisely, we use the equivalent expression of the Ricci identity $\nabla^b \nabla_b \xi_a= -R_{ab}\xi^b$ for the Killing vector
\be
\mathrm{d} * \mathrm{d} \xi = * K ,~~~ K_a \equiv 2 R_{ab} \xi^b=2 (m^2 \phi^2-2) \xi_a\,.
\ee
Integrating the 2-form $* K$ on the timelike surface $\Sigma_h^i$ of $t=t_0$ from $z_h$ to $z_i$, and using Stokes' theorem, we have
\be
\label{eq:stokes}
\int_{\Sigma_h^i} * K= \int_{\Sigma_h^i} {\mathrm{d} * \mathrm{d} \xi} =\int_{\partial \Sigma_h^i} * \mathrm{d} \xi\,
\ee
with nonzero contributions to the above equation
\be
\begin{aligned}
      (* K)_{xz}&=2 (m^2 \phi^2-2)\sqrt{-g},~~~
       (*\mathrm{d} \xi)_x&= -e^{-\chi/2}f'/z\,.
\end{aligned}
\ee 
So the left and right sides of \eqref{eq:stokes} can be calculated directly. The final result is
\be
\begin{aligned}
       \label{int-res}
\int_{\Sigma_h^i} \,\mathrm{d}x\mathrm{d}z\,\sqrt{-g} \,2 (m^2 \phi^2-2) &= -2\pi\left( \frac{\lvert f'(z_h) \rvert e^{-\frac{\chi_h}{2}}}{z_h} + \frac{\lvert f'(z_i) \rvert e^{-\frac{\chi_i}{2}}}{z_i} \right)
\end{aligned}
\ee
which can be simplified to 
\be
\begin{aligned}
    T_h S_h + T_iS_i= 2\pi \int_{z_h}^{z_i}
    \frac{e^{-\chi/2}}{z^3}\,(4-2 m^2 \phi^2)
    \,{\mathrm{d}z}\,.
\end{aligned}
\ee
It has the same form with \eqref{eq:ts}, so it also fails to prove no inner horizon.

In the following, we provide an argument to show the nonexistence of the inner horizon for black holes of case II in Sec. \ref{subsec:cII}. We consider the probe limit of the system, i.e. a probe complex scalar field in the BTZ black hole background, to study the stability of inner horizon following \cite{Balasubramanian:2004zu,Dias:2019ery}. We rewrite some useful formulas here for convenience. The metric of BTZ black hole is \eqref{eq:btz}, i.e. 
\begin{align}
\label{eq:btz-a2}
\begin{split}
    ds^2 &= \frac{1}{z^2} \left( -f dt^2 + \frac{dz^2}{f} + (N dt + dx)^2 \right )\,,\\
    f &= \frac{(z^2-z_h^2)(z^2-z_i^2)}{z_h^2z_i^2}\,, ~
    ~~N= -\frac{z^2}{z_hz_i}\,,
\end{split}
\end{align}
where $z_h$ is the outer horizon and $z_i$ is the inner horizon. 
The surface gravity $\kappa$ and the angular velocity $\Omega$ of the event horizons are
\begin{align}
    \kappa_{h}= \frac{z_i^2-z_h^2}{z_h z_i^2}\,, ~~~\Omega_h= \frac{z_h}{z_i}\,, 
\end{align}
and similarly for the inner horizon
\begin{align}
    \kappa_{i}= \frac{z_i^2-z_h^2}{z_h^2 z_i}\,, ~~~\Omega_i= \frac{z_i}{z_h}\,.
\end{align}
We will consider the non-extremal BTZ black hole. 

In the probe limit, the complex scalar field satisfy the following Klein-Gordon equation 
\begin{align}\label{eq:kg}
    \nabla^a\nabla_a\varphi-m^2\varphi=0\,.
\end{align}
For convenience, we introduce a new coordinate
\be
\tilde z=\frac{z_h^2(z_i^2-z^2)}{z^2(z_i^2-z_h^2)}\,. 
\ee
In this new coordinate, the inner horizon of BTZ is at $\tilde z=0$, the outer event horizon is at $\tilde z=1$ and the AdS boundary is at $\tilde z=\infty$. 

We consider the profile of the scalar field 
\begin{align}
\begin{split}
    \varphi &= e^{-i \omega t+ i n x}\, \phi(\tilde z)
    =e^{-i \omega t+ i n x} \,\tilde{z}^{-i \frac{\omega-n \Omega_i}{2\kappa_i}}(1-\tilde z)^{-i \frac{\omega-n \Omega_h}{2\kappa_h}} \, F(\tilde z)\,.
\end{split}
\end{align}
Then $F$ should satisfy a hypergeometric equation
\begin{align}\label{eq:hyper}
    \tilde z(1-\tilde z)F''(\tilde z)+ (c-\tilde z(a+b+1))F'(\tilde z)-ab F(\tilde z)=0
\end{align}
with
\begin{align}
    \begin{split}
        a &=\frac{1}{2}\left( \Delta -i \frac{\omega-n \Omega_i}{\kappa_i} -i \frac{\omega-n \Omega_h}{\kappa_h} \right)\,,\\ 
        b &=\frac{1}{2}\left( 2- \Delta -i \frac{\omega-n \Omega_i}{\kappa_i} -i \frac{\omega-n \Omega_h}{\kappa_h} \right)\,,\\ 
        c&= 1-i \frac{\omega-n \Omega_i}{\kappa_i}\,.
    \end{split}
\end{align}
Here $\Delta$ is the conformal dimension of dual scalar operator which satisfies  
\begin{align}
    m^2=\Delta(\Delta-2)\,.
\end{align}
Near the outer event horizon $\tilde z =1$, the solution of \eqref{eq:hyper} is 
\begin{align}
    F= c_1\ _2F_1 \big( a,b,a+b-c+1,1-\tilde z \big)  + c_2\ (1-\tilde z)^{(c-a-b)}\ _2F_1 \big( c-b,c-a,c-a-b+1,1-\tilde z \big)
\end{align}
where $c_1, c_2$ are constants of integration. We consider stationary scalar clouds which satisfy $\omega=n \Omega_h$ \cite{Hod:2012px}. This condition is essentially \eqref{eq:syncon} and similar consideration can also be seen in the four dimensional case \cite{Brihaye:2016vkv}.  With the ingoing condition at the outer horizon, the scalar field should take the form
\begin{align}
    \phi= \phi_h~ \tilde{z}^{-i \frac{\omega-n \Omega_i}{2\kappa_i}} \ _2F_1 \big( a,b,a+b-c+1,1-\tilde z \big)
\end{align}
We will analyze its behavior near the inner horizon for this solution. Near the inner horizon $\tilde{z}\to 0$, the scalar field has the form
\begin{align}
    \phi \sim \phi_h\,\frac{ \pi \csc(c\pi)}{\Gamma(a)\Gamma(b)\Gamma(2-c)} \cos\left(\frac{\omega-n \Omega_i}{2\kappa_i} \log(\tilde{z})\right)
\end{align}
which oscillates dramatically with constant amplitude and the frequency of the oscillation increases when approaching the inner horizon. This fact indicates that the inner horizon will be backreacted and destroyed due to the scalar filed. 

Moreover, we can study the behavior of energy flux at the inner horizon. The energy momentum tensor of scalar field is
\begin{align}
    T_{ab}=\frac{1}{2}  (\partial_a \varphi \partial_b \varphi^* + \partial_a \varphi^* \partial_b \varphi) - \frac{1}{2}g_{ab} \left(  \partial_c \varphi \partial^c \varphi^* + m^2 \varphi\varphi^* \right)
\end{align}
We first define the tortoise coordinate for \eqref{eq:btz-a2}, 
\begin{align}
    \frac{dr_*}{dz}=-\frac{1}{f}\,,~~~ r_*=-\frac{1}{2\kappa_i} \log\left[ \frac{z_i-z}{z_i+z} \left(\frac{z+z_h}{z-z_h}\right)^{z_i/z_h} \right]
\end{align}
the inner horizon and event horizon correspond to $r_*=\infty$ and $r_*=-\infty$ respectively. Then we can define the advanced and retarded time coordinates
\begin{align}
    u=t-r_*\,,~~~ v=t+r_*\,.
\end{align}
Finally we define the Kruskal coordinate near the inner horizon
\begin{align}
    U=-e^{\kappa_i u}\,,~~~  V=-e^{-\kappa_i v}\,, ~~~ \tilde{x}= x-\Omega_i\, t
\end{align}
where the inner horizon is located at $V=0$.

The energy flux at inner horizon 
\begin{align}
    T_{VV}&= \frac{\partial x^a}{\partial V} \frac{\partial x^b}{\partial V} T_{ab}\nonumber\\
    &=\frac{1}{4\kappa_i^2 V^2} \left( T_{tt}+2\Omega T_{tx} +\Omega^2 T_{xx} + f^2 \left(\frac{\partial \tilde{z}}{\partial z}\right)^2 T_{\tilde{z}\tilde{z}} \right)\,,
\end{align}
where $x^a=\{t, x, z\}$. 
For the solution we considered, we have 
$T_{VV} \sim 1/V^2$ near the inner horizon $V\to 0$. This divergent energy momentum tensor will contribute to the Einstein's equation and destroy the inner horizon. Therefore, the inner horizon of the BTZ black hole in the probe limit is not stable. Similar results has been studied in \cite{Balasubramanian:2004zu, Dias:2019ery, Emparan:2020rnp, Kolanowski:2023hvh}.



\end{document}